\documentclass[12pt]{article}
\usepackage{epsf}
\input epsf.tex
\usepackage{amssymb}
\usepackage{amsmath}

\newcommand{\bmat}{\left(\begin{array}}
\newcommand{\emat}{\end{array}\right)}

\def\yzero{\smash{\hbox{$y\kern-4pt\raise1pt\hbox{${}^\circ$}$}}}

\def\a{\alpha}
\def\b{\beta}
\def\g{\gamma}
\def\G{\Gamma}
\def\d{\delta}
\def\e{\epsilon}

\def\um{\frac{1}{2}}

\def\Om{\Omega}
\def\om{\omega}

\def\l{\lambda}
\def\L{\Lambda}

\def\-{\hphantom{-}}

\def\s2{\frac{1}{2}}

\def\oh{\frac{1}{2}}

\def\uq{\frac{1}{4}}
\def\IT{\bf T}
\def\beq{\begin{equation}}
\def\eeq{\end{equation}}
\def\beqa{\begin{eqnarray}}
\def\eeqa{\end{eqnarray}}

\def\Tr{{\rm Tr \,}}

\def\IF{\relax{\rm I\kern-.18em F}}
\def\II{\relax{\rm I\kern-.18em I}}
\def\IP{\relax{\rm I\kern-.18em P}}
\def\IC{\relax\hbox{\kern.25em$\inbar\kern-.3em{\rm C}$}}
\def\IR{\relax{\rm I\kern-.18em R}}

\def\Dsl{\,\raise.15ex\hbox{/}\mkern-13.5mu D} 
\def\IZ{\bf Z}
\def\IC{\bf C}

\def\D{\Delta}
\newcommand{\drawsquare}[2]{\hbox{%
\rule{#2pt}{#1pt}\hskip-#2pt
\rule{#1pt}{#2pt}\hskip-#1pt
\rule[#1pt]{#1pt}{#2pt}}\rule[#1pt]{#2pt}{#2pt}\hskip-#2pt
\rule{#2pt}{#1pt}}
\newcommand{\Ysymm}{\raisebox{-.5pt}{\drawsquare{6.5}{0.4}}\hskip-0.4pt%
        \raisebox{-.5pt}{\drawsquare{6.5}{0.4}}}
\newcommand{\Yasymm}{\raisebox{-3.5pt}{\drawsquare{6.5}{0.4}}\hskip-6.9pt%
        \raisebox{3pt}{\drawsquare{6.5}{0.4}}}

\catcode`\@=11
\newdimen\@rotdimen
\newbox\@rotbox
\def\@vspec#1{\special{ps:#1}}
\def\@rotstart#1{\@vspec{gsave currentpoint currentpoint translate
   #1 neg exch neg exch translate}}
\def\@rotfinish{\@vspec{currentpoint grestore moveto}}
\def\@rotr#1{\@rotdimen=\ht#1\advance\@rotdimen by\dp#1%
   \hbox to\@rotdimen{\hskip\ht#1\vbox to\wd#1{\@rotstart{90 rotate}%
   \box#1\vss}\hss}\@rotfinish}
\def\@rotl#1{\@rotdimen=\ht#1\advance\@rotdimen by\dp#1%
   \hbox to\@rotdimen{\vbox to\wd#1{\vskip\wd#1\@rotstart{270 rotate}
   \box#1\vss}\hss}\@rotfinish}%
\def\@rotu#1{\@rotdimen=\ht#1\advance\@rotdimen by\dp#1%
   \hbox to\wd#1{\hskip\wd#1\vbox to\@rotdimen{\vskip\@rotdimen
   \@rotstart{-1 dup scale}\box#1\vss}\hss}\@rotfinish}%
\def\@rotf#1{\hbox to\wd#1{\hskip\wd#1\@rotstart{-1 1 scale}%
   \box#1\hss}\@rotfinish}%
\def\rotate{\@ifnextchar[{\@rotate}{\@rotate[l]}}
\def\@rotate[#1]#2{\setbox\@rotbox=\hbox{#2}\@nameuse{@rot#1}\@rotbox}
\catcode`\@=12

\begin{document}
\makeatletter
\@addtoreset{equation}{section}
\makeatother
\renewcommand{\theequation}{\thesection.\arabic{equation}}
\pagestyle{empty}
\rightline{CAB-IB/2601006} \rightline{\tt hep-th/0603217}
\vspace{0.5cm}
\begin{center}
{\LARGE {On Susy Standard-like models from orbifolds of $D=6$ Gepner orientifolds \\[10mm]
} }{\large {G. Aldazabal$^{1,2}$, E. Andr\'es$^{1}$, J. E.
Juknevich$^3 $
\\[2mm]
} }{\small {\ $^1$Instituto Balseiro, CNEA, Centro At{\'o}mico Bariloche,\\[%
-0.3em]
8400 S.C. de Bariloche, and \ $^2$CONICET, Argentina.\\[3mm]{$^3$Department of Physics and Astronomy, Rutgers University, \\
Piscataway, NJ
08855 USA}

} {\bf Abstract} \\[7mm]
}
\begin{minipage}[h]{14.0cm}
As a further elaboration of the proposal of Ref.\cite{aaj} we address the construction of Standard-like models from
configurations of stacks of orientifold planes and D-branes on an
 internal space with the structure ${( Gepner \,\, model)^{c=6} \times \IT^{2}}/\IZ_N$.

As a first step, the construction of $D=6$ Type II B orientifolds on Gepner points, in the diagonal invariant case and for  both, odd and even, affine levels is  discussed. We build up the explicit expressions for B-type boundary states and crosscaps and obtain the  amplitudes among them. From such amplitudes  we read the corresponding  spectra and the tadpole cancellation equations.

 Further compactification on a $\IT^{2}$ torus, by simultaneously orbifolding the Gepner and the torus internal sectors, is performed.
The embedding of the orbifold action in the brane sector breaks the original gauge groups and leads to ${\cal N}=1$ supersymmetric chiral spectra. Whenever even orbifold action on the torus is considered, new  branes, with worldvolume transverse to torus coordinates, must be included.
 The detailed  rules for obtaining the $D=4$ model spectra and tadpole equations are shown.
As an illustration we present a 3 generations Left-Right  symmetric model that can be further broken to a MSSM model.

\end{minipage}
\end{center}
\newpage \setcounter{page}{1} \pagestyle{plain}
%
\setcounter{footnote}{0}
\section{Introduction}
The quest of the Standard Model like vacua, from open string interacting conformal field theories, received considerable attention in last years.
In particular, much progress has been achieved in the context of orientifolds of Type II string compactified on Gepner models
\cite{bw9806,gm0306,aaln,blumen0310,bw0401,bhhw0401,bw0403,dhs0403,dhs0411}.
Gepner models \cite{gepnerlect} are special points of Calabi Yau manifolds, at string scale,  that allow for a description in terms of an exactly solvable rational CFT.
First string model building on rational conformal field theories was performed in heterotic string theories in the middle 80' s \cite{fkss:1989,rcftheterotic}.
First preliminary   studies of Type II orientifolds on  Gepner points were presented in \cite{angelantonj} for  six dimensions, and in \cite{bw9806} for $D=4$ dimensions.

Recent studies of open strings models on Gepner points have been based on two alternative (but equivalent) descriptions, namely, the partition function approach or the boundary state approach (see for  instance \cite{asopen} for a review). 

In the partition function approach  consistent Type II orientifold partition functions are  built up.  Once Klein-bottle closed string partition function is identified,  M\"obius strip and cylinder amplitudes are included for consistency. The string spectrum can, therefore,  be read out from them.
Consistency implies factorization, tadpole cancellation and integer particle states multiplicities (see, for instance \cite{aaln} for details).
On the other hand, one loop open string amplitudes can be expressed in terms of closed strings propagating among boundary and crosscap states.
Once such states are identified, tadpole cancellation conditions and spectrum can be found in terms of  the quantum numbers labeling those states (see for instance, \cite{ReS,bhhw0401,bw0403}).
Either approach has lead to considerable progress.
The rules for computing spectra and the tadpole cancellation equations  have been derived for generic situations. Moreover, connections with a geometric large volume descriptions were established \cite{ABbranes,bhhw0401}.
Nevertheless,  even if concise and rather simple generic  expressions can be obtained, the computation of spectra for specific models can become rather cumbersome due to the, generically, huge number  of open string states involved. Only solving the tadpole consistency equations can represent a difficult task even for a fast computer. Therefore a systematics is needed in order to be able to extract any useful information.
In this sense a remarkable computing search for models with Sandard like model spectra was performed in \cite{dhs0403,dhs0411, Anastasopoulos:2006da} by restricting the scan to  four stacks of SM branes, by following the ideas advanced in \cite{imr} in the context of intersecting brane models \cite{intersec} on toroidal like manifolds. In fact, thousands of SM like models were found. It is worth mentioning that even the simplest of these models requires to introduce a huge number of projections and to solve several tadpole equations.

In Ref. \cite{aaj} a hybrid Type IIB orientifold construction was proposed where the internal sector is built up from a Gepner sector times a torus. By choosing a torus invariant under some of the known $\IZ_N$ phase symmetries of Gepner models, an orbifold by such symmetries was then  performed. Thus, schematically, in $D=4+2n$  the internal sector is given by ${( Gepner \,\, model)^{c_{int}=9-3n} \times \IT^{2n}}/\IZ_N$ (where $c_{int}$ is the internal central charge).
 The orbifold action is simultaneously embedded as a twist on Chan Paton factors on the open string sector resulting in a breaking of the starting $D$ dimensional Gepner orientifold gauge groups. In particular, such constructions lead to ${\cal N}=1$ $D=4$ chiral models. Illustrating examples were presented for odd affine Kac-moody levels.
 Hybrid Gepner-torus models  have some interesting features.  An important, practical,  observation \cite{aaj} is that the number of Gepner models (see \cite{fkss:1989}) involved, 3 in $D=8$ or 16 in $D=6$, is remarkably lower than the 168 models in $D=4$ (without including moddings) and so it is the number of internal states.
Also, many features can be studied analytically without the need of computers.
From the phenomenological point of view, the possibility of having large extra dimensions, in the torus directions, could be an appealing feature allowing for some control over the string scale.

In this note we  elaborate on this proposal of hybrid models. We concentrate on  $D=6$ Gepner models, with diagonal invariant couplings, and extend the results of \cite{aaj} to include  both, odd and even, affine Kac-Moody levels.
$D=6$ models present particular features that make them interesting \textit{per se} (see for instance \cite{fkllsw}). Moreover, due to the presence of potential gravitational and gauge anomalies these models  are particularly useful to test the consistency of the construction.

We build up the explicit expressions for B-type boundary states and crosscaps and obtain the corresponding amplitudes for strings propagating among them. From such expressions we read the tadpole cancellation equations and the rules for reading  the  spectra.
An explicit example (the $6620$ model) is developed in detail. Results for the 16 six dimensional models are summarized in \cite{aajd6gepners}. As far as we are aware of, besides the first examples of Gepner orientifolds  in Ref.\cite{angelantonj}, only some other $D=6$ spare examples (see for instance \cite{aaln,aaj,Braun:2005eg}) appear in the accessible literature.

Following \cite{aaj} we further compactify on a $\IT^{2}$ torus by simultaneously orbifolding the Gepner and the torus internal sectors and by  embedding  the orbifold action on  the brane sector.
 Interestingly enough, whenever even orbifold action on the torus is considered, new  branes, with worldvolume transverse to torus coordinates, must be generically included for consistency requirements\footnote{This is in fact expected. It parallels the inclusion of a 55 sector, besides a 99  brane sector, when even twists are present in $\IZ_{2N}$ orientifold compactifications.}. Detailed  rules for obtaining the $D=4$ model spectra and tadpole equations are shown.

As an illustration we show how to obtain a 3 generations Left-Right  symmetric model (which can be further broken into a MSSM model) from a $Z_4$ orbifold of the, $D=6$,  $6620$ diagonal Gepner model times a torus..

The article is organized as follows. Section 2 contains a generic  introduction to Type IIB orientifolds, crosscaps and boundary states.
 In Section 3 orientifold of  $D=6$ Gepner models are discussed and crosscap and boundary states are constructed. The  rules for computing the spectra and tadpole cancellation equations are derived. The explicit example $6620$ is discussed in detail. Section 4 provides a generic discussion of hybrid compactifications
${( Gepner \,\, model)^{c_{int}=6} \times \IT^{2}}/\IZ_N$. In  Section 5 we construct a MSSM like example as a $\IZ_4 $ modding of $6620\times \IT^{2}$ and discuss some, generic,  phenomenological features.
Computation details are collected in the Appendices.

\section{Type II orientifolds, crosscaps and boundary states}

In this section we   briefly review the basic steps in the construction of orientifold models.
Essentially, an orientifold model is obtained by dividing out the orientation reversal symmetry of Type II string theory (see for instance \cite{asopen, aaln}).
Schematically, Type IIB torus partition function is defined as
\begin{equation}\label{2bpf}
{\cal Z}_{T} (\tau, {\bar \tau})=   \sum_{a,b} \chi_a(\tau )
{\cal N}^{ab} {\bar \chi}_b({\bar \tau })
\end{equation}
where the characters $\chi_a(\tau )= \Tr_{{\cal H}a}
q^{L_0-\frac{c}{24} }$, with $q= e^{2i\pi \tau}$, span a representation
of the modular group of the torus
generated by {\sf S}: $\tau \to -\frac1{\tau}$ and {\sf T}: $\tau
\to\tau+1$ transformations.
${\cal H}_a$ is the Hilbert space of a conformal field theory with central
charge
$c=15$, 
generated from
a conformal primary state $\phi _a$
(similarly for the right moving algebra).
In particular $\chi_a( -\frac1{\tau} )= S_{aa'}\chi_{a'}(\tau )$
and modular invariance require 
$ S {\cal N} S^{-1}= {\cal N}$ (for left -right symmetric theories ${\cal N}^{ab}={\cal N}^{ba}$).
Generically, the characters can be split into a spacetime piece,
contributing with
$c_{st}= {\bar c }_{st}= {\frac32} D$ and an internal sector with
$c_{int} = {\bar c}_{int}= {\frac32}(10-D)$.

Let $\Omega $ be the reversing order (orientifolding)  operator permuting
right and left movers.
Modding by  order reversal symmetry is then
implemented by introducing the projection operator
$\frac12 (1+\Omega)$ into the torus partition function.
The resulting vacuum amplitude reads
\begin{equation}\label{clomega}
{\cal Z}_{\Omega} (\tau, {\bar \tau})= {\cal Z}_{T} (\tau, {\bar \tau})
+ {\cal Z}_{K} (\tau- {\bar \tau}) .
\end{equation}
 The first term is just the symmetrization
(or anti-symmetrization in
case states anticommute) of left and right sector contributions
indicating that two states differing in a left-right ordering must be
counted only once.
The second term is the
Klein bottle contribution and takes into account states that are
exactly the same in both sectors.
In such case, the operator
$e^{2i \pi \tau {L_0}}
e^{-2i \pi {\bar \tau}{{\bar L}_0}}$,
when acting on the same states, becomes
$e^{2i \pi 2it_K {L_0}}$ with $\tau -{\bar {\tau}}= 2it_K$ and thus
\begin{equation}\label{kbd}
{\cal Z}_{K} (2it_K) = \frac 12 \sum_{a} {\cal K}^{a}  \chi_a(2it_K) ,
\end{equation}
where $|{\cal K}^{a}|={\cal N}^{aa}$.
The Klein bottle
amplitude in the {\it transverse channel} is obtained by performing
an {\sf S} modular transformation
such that
\begin{equation}\label{kbo}
\tilde {\cal Z}_{K} (i l)= \frac 12 \sum_{a} O^2_a  \chi_a(il )
\end{equation}
with $l=\frac1{2t_K}$ and
\beq
(O^a)^2= 2^D {\cal K}^{b}S_{ba}
\eeq
This notation for the closed channel coefficients highlights the fact
that the Klein bottle
 transverse channel represents a closed string propagating
between two
crosscaps (orientifold planes) states. Namely, a quantum state
$|C\rangle $,  describing the crosscap can be found such that the KB amplitude can be expressed as
\begin{equation}
\tilde {\cal Z}_{K} (i l)=\frac12 \langle C |q^{\frac12 H_{cl}}|C\rangle .
\end{equation}
with $H_{cl}=L_0-\tilde L_0-\frac c{12}$.

Indeed, crosscap states can be formally expanded in terms of Ishibashi states \cite{ishi,ooy} such that
\begin{equation}
|C\rangle= O_a |a\rangle\rangle _C
\end{equation}
with
\begin{equation}
_C\langle\langle b |q^{\frac12 H_{cl}}|a\rangle\rangle_C=\delta_{a,b}\chi^{a}({\tilde q})~,
\end{equation}
and  ${\tilde q}=e^{2i\pi l}$.

When integrated over
the tube length, such amplitude leads, for massless states, to tadpole like divergences. In particular, for RR massless states, such tadpoles must be cancelled for the theory to be consistent. Notice that, for such fields, $O_a$ represents the charge of the orientifold plane (crosscap) under them and, therefore, inclusion of an open string sector with D-branes carrying $-O_a$ RR
 charge provides a way for having a consistent theory \cite{pol,gp, pchj} with net vanishing charge.

Therefore, we  introduce  stacks  of  boundary states $|\alpha\rangle$ (referred to as ``brane-$\alpha$") 
\begin{equation}
|\alpha\rangle= \sum_{a} D_{\alpha}^a |a\rangle\rangle_B
\end{equation}

such  that the  amplitude, describing propagation of strings between "intersecting"    stacks ${ \alpha}$ and ${ \beta}$  can be written as
\begin{equation}
\tilde {\cal Z}_{{ \beta}, { \alpha}}(i l)=
\langle \beta |q^{\frac12 H_{cl}}|\alpha\rangle=
\sum_{a} D_{\alpha}^a D_{\beta}^a\chi^{a}(l)=\sum_{b} C_{{ \beta}, { \alpha}}^b\chi^{b}(t/2)
\label{alfabeta}
\end{equation}
where in the last step we have perform an $S$ modular transformation to direct channel. 

Here
\begin{equation}
C_{{ \beta}, { \alpha}}^b=\sum_{a} D_{\beta}^a D_{\alpha}^a S_{ab}
\end{equation}
is  the multiplicity of open string states contained in
$ \chi_a$.
Namely, it counts  open string sector states  of the form
\begin{equation}
|\Phi_a ; {\beta},{\alpha}\rangle \
\end{equation}
where $\Phi_a$ is a world sheet  conformal field and
${\alpha},{\beta}$  label the
type of ``branes" where the string endpoints
must be attached to.  $C_{ \beta, { \alpha}}^a$ are positive integers (actually $C_{ \beta, { \alpha}}^a= 0,1,2$ \cite{aaln}) generated when the trace over open states $|\Phi_a;{\beta},{\alpha}\rangle$ is computed.

The full cylinder partition function is obtained when summing over all possible stacks  of $n_{ \alpha}$ branes, namely
\begin{equation}
{\tilde {\cal Z}}_C (i l)=\sum_{a} {D_a} ^2\chi^{a}(l)
\end{equation}
with $D^a=\sum_{\alpha} n_{ \alpha} D_{\alpha}^a$.

In a similar manner, strings propagating between  branes and  orbifold planes give rise to  strip amplitude
\footnote{In order to obtain the above expressions  we have used that
\begin{eqnarray}
_B\langle\langle a|q^{\frac12 H_{cl}}|b\rangle\rangle_B&=&\delta_{a,b}\chi^{a}(l)~,\\
_C\langle\langle b |q^{\frac12 H_{cl}}|a\rangle\rangle_B=\delta_{a,b}\chi^{a}({\tilde q})&=&\delta_{a,b}{\hat\chi}^{a}(l+\frac12)~,
\end{eqnarray}
where
${\hat
\chi}_a(il+\oh )=e^{i\pi
(h_a-c/24)} \chi_a (it_{M}+\frac{1}{2})$ is real.}

 \begin{equation}
{\tilde {\cal Z}}_M (il)=2 D_a O_a{\hat\chi}^{a}(l+\frac12)
\end{equation}

By modular transforming to direct channel we obtain multiplicities of open string states between a brane and its orientifold image
\beq
O_a ( n_{ \alpha} D_{\alpha}^b)P_{ba}= {\cal M}_{a} = M_{\alpha}^bn_{ \alpha}
\eeq
where we have used the fact that  characters in the direct and transverse channels of the M\"obius strip
are related by the transformation \cite{bs1}
{\sf P}: $it_M+\frac{1}{2}
\to \frac{i}{4t_M}+\frac{1}{2}$
generated
from the modular transformations {\sf S} and {\sf T} as
${\rm \sf P}={\rm {\sf TST}}^{2}{\sf S}$.

For indices $a$ representing massless RR fields  $ D_a$ is the D-brane RR charge. Therefore  zero net RR charge  requires the
\begin{equation}
O_a +n_{ \alpha} D_{\alpha}^a                                                                                                                                                                                                                                         =0 .
\label{tadpolecg}
\end{equation}
tadpole cancellation equations.

\section{Orientifolds of $D=6$ Gepner models}
In this section we briefly summarize the main ingredients involved  in the
construction Gepner model orientifolds  in six spacetime
dimensions. We refer the reader to the appendices and references for a survey of the details.
In Gepner models \cite{gepnerlect}, in $D$ space time dimensions, the internal sector is given by a tensor product of $r$
copies of $N=2$ superconformal minimal models with levels $k_j$, $j=1,...,r$
and central charge
\begin{equation}
c_{j} = \frac{3{k_j}}{{k_j}+2} \quad , \quad {k_j}=1,2,...  \label{mm}
\end{equation}
such that internal central charge $c_{int}=\sum _{j=1}^r c_{j}^{int}=12-3(D-2)/2$.

 Unitary representations of $N=2$ minimal models  are
encoded in primary fields   labelled by three integers $(l,m,s)$ such that $l=0,1,...,k$; $%
l+m+s=0$ mod 2. These fields belong to the NS or R sector when $l+m$ is
even or
odd respectively\footnote{Recall that two representations labelled by $(l^{\prime},m^{\prime},s^{\prime})$ and $%
(l,m,s)$ are equivalent if $ l^{\prime}=l$ and $m^{\prime}= m ~{\rm mod} ~ 2(k+2)$  and $s^{\prime}=s ~ {\rm mod}~4$ or $(l^{\prime},m^{\prime},s^{\prime})=(k-l,m+k+2,s+2)$.}.
Spacetime supersymmetry and modular invariance are implemented by keeping in the spectrum only states for which the total $U(1)$ charge is an odd integer.

The primary field information of the complete theory can be conveniently encoded in the vectors $\lambda $ and $\mu $ defined in appendix B. Thus, the index $a$ in the previous section amounts here for $a=(\lambda,\mu )$ in Gepner's case.

In six dimensions $c_{int}=6$ and  16 different possible Gepner models exist,  which are associated to $K3$ surfaces \cite{fkss:1989,Braun:2005eg}. Namely,

\begin{table}[htbp]
  \centering
  \begin{tabular}{llll}
    $\overline{k}=(1,1,1,1,1,1)$, &
    $\overline{k}=(0,1,1,1,1,4)$, &
    $\overline{k}=(2,2,2,2)$,   &
    $\overline{k}=(1,2,2,4)$,   \\
    $\overline{k}=(1,1,4,4)$,   &
    $\overline{k}=(1,1,2,10)$,  &
    $\overline{k}=(0,4,4,4)$,   &
    $\overline{k}=(0,3,3,8)$,   \\
    $\overline{k}=(0,2,6,6)$,   &
    $\overline{k}=(0,2,4,10)$,  &
    $\overline{k}=(0,2,3,18)$,  &
    $\overline{k}=(0,1,10,10)$, \\
    $\overline{k}=(0,1,8,13)$,  &
    $\overline{k}=(0,1,7,16)$,  &
    $\overline{k}=(0,1,6,22)$,  &
    $\overline{k}=(0,1,5,40)$
  \end{tabular}
  \caption{Gepner models associated to $K3$}.
  \label{tab:K3}
\end{table}

Notice that, in some cases, $k=0$ blocks have been added. Even if such terms are irrelevant in a closed string theory (for instance the central charge remains invariant),  they have been shown to have a non trivial (K-theory) effect when open string sector is included. In fact, an even (odd) number of internal minimal blocks is required (see for instance \cite{bhhw0401,fkllsw}) in $D=6$ ($D=4$) for consistency\footnote{Although in \cite{Braun:2005eg} the extra case $(0,1,1,1,1,1,1)$ is suggested to be associated to a different K3 surface  we have not explored this possibility. Also the $(0,1,1,1,2,2)$ models correspond to tori surfaces.}.

Actually, for the sake of simplicity we will consider the case where the
internal sector is a tensor product of $r = 6 $ conformal blocks.
 This will allow us to simultaneously consider
cases with 3, 4, 5, and 6 conformal blocks such as $(6)^2(2)$,
$(2)^4$, $1^4$ or $(1)^6$ by adding, if necessary, conformal blocks with
level $k=0$.
On the other hand,  as it is shown in ~\cite{bhhw0401},
 the formulae for the total crosscap states contain the
 sign factor $(-1)^{\mu}$ where the parameter $\mu$ is given by

 \beq
\mu = \sum_{i=1}^r \left ( 1- \frac{1}{k_i+2} \right
)=\frac{r+2}{2} \eeq When $r=6$ there are no extra signs due to
$(-1)^{\mu}$ and hence the expressions for the crosscap, that we will derive
below, become somewhat simpler.

The Klein bottle amplitude is determined from that of the torus up
to signs representing different ways of ``dressing'' the
world-sheet parity $\Omega$. We  will denote the dressed parity  (we closely follow the notations in references \cite{aaj,blumen0310}) as
$\Omega_{\Delta, \omega_j}^{B}$, where $B$ means we are dealing
with $B$-type orientifolds and $\Delta$, $\om_j$ label the quantum
and phase symmetries respectively.

Recall that in four-dimensional Gepner models the B-parity is
related to the $A$-parity $\Om_{\om, \Delta_i}^A$ via the
Green/Plesser~\cite{greenplesser}  mirror construction

\beqa \Om_{\Delta, \om_i}^{B} ~~ &\leftrightarrow& ~~\Om_{\om,
\Delta_i}^{A}  \\
\Delta = H \sum_{i=1}^r\frac{\Delta_i}{k_i+2} && \om=\sum_{i=1}^r
\om_i \eeqa

where $H = lcm \{ k_i+2\}$.

Following  ~\cite{blumen0310} we define an orientifold projection
$\Om_{\Delta, \omega_j}$ by including the sign factors \beq
(-1)^{\sum\om_j \Lambda_0} (-1)^{\D b/H}\prod_j (-1)^{\om_j m_j} \label{signos}
\eeq where $\Delta, \om_j =0, 1$. These signs or parity dressings
are chosen so that they preserve supersymmetry. By introducing these signs and by computing the trace in (\ref{kb}) we are lead to
\beqa
Z^B_{K \om_j,\D} &=&\frac{4}{(8\pi^2 \alpha')^3} \int_0^\infty
{\frac{dt} {t^3}}\frac{1}{2^{r+1}}\frac{1}{\eta(2it)^3}
\sum_{\lambda, \mu}^{\beta} \sum_{\eta_1,\cdots,\eta_r=0}^1 \sum_{b=0}^{\frac{K}{2}-1}(-1)^{\Lambda_0}(-1)^{\sum\om_j \Lambda_0}\\\nonumber
&&(-1)^{\D b/H}\prod_j (-1)^{\om_j m_j}\left (
\prod_{k<l}(-1)^{\eta_k
\eta_l} \right) \\\nonumber
&&\prod_j \d_{b , \,\eta_j \,(k_j+2)}^{(2k_j+4)} \prod_j
\d_{\eta_j l_j, \,\eta_j \frac{k_j}{2}} \chi_{\mu}^{\lambda}
\eeqa

where  $K={\rm lcm}(4,2k_j+4)$ (see appendix B for notation).

The factor $\prod_{k<l}(-1)^{\eta_k \eta_l}$ is introduced for
convenience and   arises naturally from the definition of the
crosscap state  below.

From the Klein bottle amplitude in the transverse channel we can
read the expression for the crosscap state up to signs that can be
fixed from the M\"obius strip amplitude. The result is that the
crosscap state is given by\footnote{For explicit expressions for modular matrices $P_{l',l}$ and $S_{l',l}$ see \cite{aaj,blumen0310}.}

\begin{eqnarray}
\label{crosscapbf}
 | C\rangle^{NS}_B &=& \frac1{\kappa_c}
   {\sum_{\lambda',\mu'}}^{ev}  \sum_{\nu_0=0}^{\frac{K}2-1 }
      \sum_{{\nu_j} , \tilde \nu_{1}=0}^1  \sum_{\epsilon_j=0}^1
   \,\, \left(\prod_{k<l} (-1)^{\nu_k\nu_l} \right) (-1)^{\nu_0} \,(-1)^{\sum_j \nu_j}
\\ \nonumber
  & &   \,
(-1)^{\sum \omega_j \Lambda'_0 /2} e^{2\pi i \nu_0 \sum
\frac{\Delta_j}{k_j+2}}\, \delta_{\Lambda'_0,2+2\nu_0 +2\tilde \nu_{1}
       +2\sum \nu_j+2\sum \om_j }^{(4)} \delta_{\Lambda'_{1},2\nu_0 +2\tilde \nu_{1}
     }^{(4)} \\\nonumber
  && \prod_{j=1}^r \Biggl(  \sigma_{(l_j',m_j',s_j')}\,
\frac{P_{l'_j,\epsilon_j k_j}}{\sqrt{S_{l'_j,0} }}  \,
    \delta_{m_j',2\nu_0+(1-\epsilon_j+\omega_j)(k_j+2)}^{(2k_j+4)}\\\nonumber
&& \delta_{s_j',2\nu_0 +2 \nu_j+2(1-\epsilon_j)}^{(4)}
(-1)^{\epsilon_j {\frac{(m'_j+s'_j)} 2} } \Biggr)
|\lambda',\mu'\rangle\rangle_c
\end{eqnarray}

The normalization is chosen so that the overlap of the crosscap
with itself yields the transverse Klein   amplitude

 \beq \label{overlap}
\widetilde{Z}_K^B= \um\int_0^\infty  {dl} \langle C|e^{-2\pi l H_{cl}} |
                C\rangle_B . \eeq

In order to cancel  tadpole-like divergences, boundary states must be introduced. We consider the
 B-type RS-boundary states~\cite{ReS}

\beqa \label{borde} |\alpha\rangle_B&=&|S_0, \tilde S_{1};(L_j,M_j,S_j)_{j=1}^r
\rangle_B=
         {\frac1{ \kappa^B_\alpha}}
       {\sum_{\lambda',\mu'}}^{\beta,b} (-1)^{\frac{\Lambda_0^2}2}\,
          e^{-i\pi{\Lambda'_0\, \frac{S_0} 2}} e^{-i\pi{\Lambda'_{1}\, \frac{\tilde S_{1}} 2}}\\ \nonumber
&&\prod_{j=1}^r \frac{S_{l'_j,L_j}}{  \sqrt{S_{l'_j,0} }}\,
 e^{i\pi\frac{m'_j\, M_j}{ k_j+2}}\,
      e^{-i\pi\frac{s'_j\, S_j}{ 2}} |\lambda',\mu'\rangle\rangle
      \eeqa

where the  "b" in the summatory implies that

\beq
m_j = b \mod{k_j+2}.
\eeq

In fact, due to supersymmetry and field identifications these B-type
boundary states only depend on ${\bf L}=(L_1,\dots,L_r)$ with  $L_i \leq
k_i/2$, $M=H \sum \frac{M_i}{k_i+2}$ and $S=\sum S_i$. However,
whenever a label $L_i$ reaches $k_i/2$,  extra copies of the gauge
field  may appear propagating on the brane world-volume. In this
case, itº is necessary to resolve the branes  into elementary branes such that
only a single gauge field is propagating on the world-volume. The
details depend on the values of $|\mathbf{S}|$ counting the number of $i$
such that $L_i=k_i/2$. It can be shown~\cite{bhhw0401} that when $|\mathbf{S}|$ is an odd
integer the elementary branes are given by

\beq |\a_{ele} \rangle_B = \frac{1}{2^{\frac{|\mathbf{S}|-1}{2}}}|\a \rangle_B
\eeq

Instead, if $|\mathbf{S}|$ is even there is an extra $\mathbf{Z}2$-valued
label $\psi$ taking values $\pm$ so that the elementary boundary
states are now labelled by $\bf{L},M,S, \psi$. The original
boundary states can be written in terms of the elementary ones as

\beq |\a \rangle_B = \frac{1}{2^{|\mathbf{S}|}} \left \{ |\a+
\rangle_B+|\a -\rangle_B \right\} \eeq where $\a$  stands for all labels
different from $\psi$. The two boundary states $|\a\pm
\rangle_B$ contain new states from the  twisted $(c,c)$ RR sector

\begin{eqnarray}
|\alpha\pm\rangle _B &=&|S_0,S_{-1};(L_j,M_j,S_j)_{j=1}^r,\pm \rangle_B \\ \nonumber
& = & {\frac{1}{ \kappa^B_\alpha}}
       {\sum_{\lambda',\mu'}}^{\beta,b} \,e^{-i\pi{\mu' \mu_B}}
           \{\otimes_{j=1}^r \frac{S_{l'_j,L_j}}{\sqrt{S_{l'_j,0}}} \, \delta_{m'_j,b}^{(k_j+2)}\,
\\ &&\pm\,\otimes_{j\in S}\delta_{l',\frac{k_j}2}  \,e^{-i\frac{\pi}2 M_j} \,
\delta_{m'_j,b+\frac12(k_j+2)}^{(k_j+2)}\otimes_{j\notin S}{\frac{S_{l'_j,L_j}}{\sqrt{S_{l'_j,0}}} \, \delta_{m'_j,b}^{(k_j+2)}\, \}
      |\lambda',\mu'\rangle\rangle }
\label{btypebs}
\end{eqnarray}
where $\mathbf{S}=\{i : l'_i=\frac{k_i}2\}$ and $\mu_B= (S_0, S_{-1};M_1,...,M_r;S_1,...,S_r)$.

Actually, the $|\a\pm
\rangle_B$ branes are necessary in order to have D-branes charged under all RR fields
in the theory. 
Geometrically the situation is as follows \cite{brunschom}. The twisted RR fields are related
to singular curves of the associated Calabi-Yau spaces. Then the elementary 
D-branes $|\a\pm\rangle_B$ can wrap on the new homological cycles arising from
the resolution of the singularities.

Let us now look at the supersymmetric spectrum in the open sector.  The boundary states (\ref{borde})  preserve
the same supersymmetry than the crosscap (\ref{crosscapbf}) if the following condition is satisfied 

\beq \label{susy_brana}
M=\D + \frac{H}{2} \sum{\om_i} \mod{2}
\eeq

The massless fields in the 6D spacetime theory are the vector field $(2,0)(0,0,0)^6$
 and the
 hypermultiplets
 $(0,0)\prod_j(l_j, l_j, 0)$ with $\sum_j \frac{l_j}{k_j +2} = 1$.
 They are contained in the cylinder and M\"obius amplitude which we
 present next.
The bosonic and massless part of the cylinder amplitude between two
D-branes $|\mathbf{L},M \rangle$ and $|\mathbf{L'},M'\rangle$ is generally
given by

\begin{equation} \label{cilindro}
\frac12\frac{1}{N_{\mathbf{S}}}\frac{1}{N_{\mathbf{S}'}}\sum_{\lambda,\mu}^{sr}\sum_{\epsilon_1,...,\epsilon_r=0}^1\left(\prod_{i=1}^{r}N_{L_j,{
{L'_j}}}^{|\epsilon_jk_j-l_j|} \right)  \delta_{\sum_i \epsilon_i =
1+\frac{s_0}{2}}^{(2)}\chi_{\mu}^{\lambda}
\end{equation}
where $N_{L_1,L_2}^{l}$ are the $SU(2)$ fusion coefficients (\ref{su2fusion})
 and 
\begin{displaymath}
N_{\mathbf{S}} =\left \{ \begin{array}{ll}
2^{|\mathbf{S}|/2} & \textrm{if } |\mathbf{S}| \textrm{ even} \\
2^{[|\mathbf{S}|-1]/2} & \textrm{if } |\mathbf{S}| \textrm{ odd} \\
\end{array}\right.
\label{ns}
\end{displaymath} 
eliminates any extra counting when some of the D-branes are short. We have already 
taken into account the condition (\ref{susy_brana}) and therefore the labels $M, M'$ do not appear
explicitely in this expression.
Besides, we
have defined an extra label  $s_0= \Lambda_0 + \Lambda_{1} \mod{4}$
(see Appendix A) taking the values $0, 2$ whenever the fields are in  the scalar and
vector representations, respectively (Note that in six  dimensional
spacetime $s_0 = 0, 2$ also for the spinor representations, so this
definition strictly makes sense when we restrict ourselves to bosonic
representations). When the amplitude  between two short-orbit
branes
 $|\mathbf{L}, \psi\rangle$ and $|\mathbf{L'}, \psi'\rangle$ such that $\mathbf{S}=\mathbf{S}'$ and $|\mathbf{S}|\in 2\IZ^{+}$ is considered,  an additional projection must be taken into account, due to the $\psi$ labels,  leading to

\begin{equation} \label{cilindroshort}
\frac{1}{2^{|\mathbf{S}|}}\sum_{\lambda,\mu}^{sr}\sum_{\epsilon_1,...,\epsilon_r=0}^1\left(\prod_{i=1}^{r}N_{L_j,{\tilde
{L_j}}}^{|\epsilon_jk_j-l_j|} \right) \delta^{(2)}_{\sum_{i \in
\mathbf{S}}^{} \epsilon_i = \frac12(1-\psi \psi')} \delta_{\sum_i
\epsilon_i = 1+\frac{s_0}{2}}^{(2)}\chi_{\mu}^{\lambda}
\end{equation}

On the other hand, the massless states in the bosonic M\"obius  strip amplitude
are given by
 \beq \label{moebius}-\frac{1}{2 N_{\mathbf{S}}}\prod_{k<l}(-1)^{\rho_k \rho_l} \delta_{\sum \rho_j + 1 +
\frac{s_0}{2},\sum \om_j}^{(2)} (-1)^{\sum \om_j \frac{s_0}{2}} e^{i
\frac{\pi}{2}\sum_j
\om_j(m_j-2L_j+\epsilon_j(k_j+2))}(-1)^{\epsilon_j} N_{L_j
L_j}^{|\epsilon_j k_j-l_j|} \hat \chi_{\mu}^{\lambda}\eeq

where $\rho_j = \frac{s_0 }{2} +1 + \epsilon_j + \sum \om_j$.

In particular,  we see the vector ($s_0=2$) has the sign

\beq -\frac{1}{N_{\mathbf{S}}}(-1)^{\sum \om_j}(-1)^{\sum \om_j L_j} \sum_{\e_1, \dots,
\epsilon_r=0}^{1} \prod_{k<l} (-1)^{\e_k\e_l} \prod (-1)^{\om_j \e_j
\frac{k_j+2}{2}} \delta_{\epsilon_jL_j,\epsilon_j\frac{k_j}{2}}
\d_{\sum \e_j , \sum \om_j} \label{signo_vector}\eeq

A plus (minus) sign indicates a symplectic (orthogonal) gauge group while a zero leads to a unitary  gauge group.
In a similar manner, the gauge group representations in which matter states transform, can be identified (an example is given in next section).

 The action of $\Om_{\D,\om_j}$ on these elementary boundary
 states can be obtained by comparing (\ref{moebiloopdr}) to the cylinder amplitude between a
D-brane $|\mathbf{L},M \rangle$ and its $\Om-$image $|\mathbf{L'},M'\rangle$. They coincide if the action is given by (see ~\cite{blumen0310} for instance)

\beq  \Om_{\D,\om_j}: | \mathbf{L},M,S \rangle \to
            |\mathbf{L},2\D-M,-S \rangle \eeq

Furthermore, consistency of (\ref{signo_vector}) with the cylinder amplitude (\ref{cilindroshort}) between a given brane with
a label $\psi$ and its  image under $\Om$ with  a label $\psi'=\Om(\psi)$ requires

 \beq
 \label{psi}\Om_{\D,\om_j}: \psi \to  (-1)^{\mu} (-1)^{|\mathbf{S}|/2} \prod_{i\in \mathbf{S}} (-1)^{\om_j  
\frac{k_j+2}{2}} \psi . \eeq

To see it we use that

\beq \frac{1}{2^{\mathbf{S}/2}} \sum_{\e_1, \dots,
\epsilon_r=0}^{1} \prod_{k<l} (-1)^{\e_k\e_l}  \delta_{\epsilon_jL_j,\epsilon_j\frac{k_j}{2}}
\d_{\sum \e_j , 0} =  \pm \um \big(1+ (-1)^{\mu} (-1)^{|\mathbf{S}|/2}  \big)  \eeq

Even though we are dealing with the case $r=6$ we have introduced the factor $(-1)^{\mu}$ to make contact with the case
$r=4$. Its origin is simple. When  we go from $r=6$ to $r=4$ subtracting  two $k=0$ factors leads to $(-1)^{|\mathbf{S}|/2} \to (-1)^{\mu}(-1)^{|\mathbf{S}|/2}$.

Thus, for instance, in the case $\D = \om_j = 0$ we find that, according to $|{\bf S}|$ values and specifying for $r=6$ , the  groups shown in Table \ref{groups} arise.

\begin{table}{}
\centering
\begin{tabular}{|l|c|c|}
\hline
$|\mathbf{S}|$ &$|B\rangle$ & Group r=6 \\
\hline \hline
0&  $*$ & $SO(N)$\\
1& $|\hat B\rangle$& SO(N)\\
2&  $|\hat B+\rangle+|\hat B-\rangle$&U(N)\\
3 & $2|\hat B\rangle$& SP(2N)\\
4 & $2(|\hat B+\rangle+|\hat B-\rangle)$&$SP(2N)\times SP(2N)$ \\
5 & $4|\hat B\rangle$&  SP(4N) \\
6 & $4(|\hat B+\rangle+|\hat B-\rangle)$
&U(4N)\\ \hline
\end{tabular}
 \label{groups}
\caption{Groups that arise from introducing a given number
of reducible D-branes are shown. Gauge symmetry is enhanced in some 
cases since  these branes can be decomposed into a set
of elementary D-branes. However,  nothing prevents us from considering simply
one copy of an elementary D-brane (and its image), thus yielding a
gauge group with unity range, ie. $U(1), Sp(2), SO(2)$.}
\end{table}
The tadpole cancellation conditions can be easily read from the
expressions for the crosscap (\ref{crosscapbf}) and boundary states (\ref{borde}). They take the
general form Tad$_{D}(\lambda,\mu)-8$Tad$_{O}(\lambda,\mu)=0$.
 For
the massless fields $(2,0)(0,0,0)^6$ and $(0,0)\prod_j(l_j, l_j, 0)$
 the NS-NS tadpoles of the orientifold plane read

\begin{eqnarray}
 \label{tadorien}
 {\rm Tad}_{O}(\lambda,\mu)_B&=&
     \sum_{\nu_0=0}^{{\frac K2}-1}
      \sum_{\epsilon_1,\ldots,\epsilon_r=0}^1
     \, \, \left(\prod_{k<l} (-1)^{\epsilon_k\epsilon_l} \right)\,(-1)^{\nu_0 \sum \epsilon_j}\,\\ \nonumber
   & & \,
\delta_{s_0/2,1+\sum \epsilon_j + \sum \om_j}^{(2)}(-1)^{\sum
\om_j (s_0/2)}(-1)^{\D_j (1-\epsilon_j)}
\\\nonumber
  && \prod_{j=1}^r \Biggl( \, {\rm sin}\left[\frac{1}{2}(l_j, \epsilon_j k_j)\right]\,
\delta^{(2)}_{l_j+(1-\epsilon_j) k_j,0} \,
    \delta_{m_j',2\nu_0+(1-\epsilon_j +\om_j)(k_j+2)}^{(2k_j+4)}\\\nonumber
&& (-1)^{\epsilon_j {\frac{m_j} 2} } \Biggr).
\label{tadod=6}
\end{eqnarray}

Also, collecting all terms from the boundary states and their
$\Omega_{{\Delta}_j,\omega,\omega_\alpha}$ images, we obtain, for their massless
tadpoles 

\beqa {\rm Tad}_{D}(\lambda,\mu)& =& \,
 \sum_{a=1}^N \frac{ N_{a}}{N_{\mathbf{S}}}\, {\rm cos}\left[\pi
     \sum_j \frac{m_j (M_j^a-\Delta_j)}{k_j+2} \right] \,
\prod_j {\rm sin}(l_j,L_j^a). \label{taddd=6}\eeqa 
These expressions
are valid up to common phases. We have also renormalized the tadpole
equations by introducing the factor $N_{\mathbf{S}}$ so that the
Chan-Paton factors $N_a$ truly represent the multiplicity of
elementary D-branes.
\subsection{$(6)^2(2)(0^3)$ model}
We exemplify the construction presented in the preceeding section for  the specific Gepner model $(6)^2(2)(0^3)$. We will later consider this example to discuss model building in four dimensions.  Results for the other six dimensional models are presented in \cite{aajd6gepners}. The allowed branes and corresponding gauge groups and matter representations living on them (see [\ref{groups}]) are given in Table \ref{662table}.
\begin{table}
\centering
\begin{tabular}{|l|c|c|rrrr|}
\hline
\# &$(L_1,L_2,L_3,L_4)$ & Group & \# (&\Ysymm&\Yasymm&)\\
\hline \hline
${\bf L_1}$& 0 0 0 $0^3$ & $Sp(N_1)$ & & 0 & 0& \\
${\bf L_2}$& 1 1 0 $0^3$& $Sp(N_2)$ && 0 & 3 &\\
${\bf L_3}$& 3 1 0 $0^3$& $Sp(N_3)\times Sp(N_3)$ & & 0 & 2  &\\
${\bf L_4}$ & 3 3 0 $0^3$& $Sp(N_4)$ & & 0 & 3 &\\
${\bf L_5}$ & 2 0 0 $0^3$ & $Sp(N_5)$ & & 0 & 2  &\\
${\bf L_6}$ & 2 2 0 $0^3$& $Sp(N_6) $ & & 1 & 7  &\\
${\bf L_7}$& 0 0 1 $0^3$& $Sp(N_7)\times Sp(N_7)$ & & 0 & 0  &\\
${\bf L_8}$& 1 1 1 $0^3$& $Sp(N_8)\times Sp(N_8)$ &&  0 & 2 &\\
${\bf L_9}$& 3 1 1  $0^3$& $Sp(N_9)$ &&  0 & 3  &\\
${\bf L_{10}}$ & 3 3 1 $0^3$& $U(N_{10})$ & &0& 6 &\\
${\bf L_{11}}$ & 2 0 1 $0^3$& $Sp(N_{11})\times Sp(N_{11})$ & & 0 & 1 &\\
${\bf L_{12} }$& 2 2 1 $0^3$& $Sp(N_{12}) \times Sp(N_{12})$ & & 0 & 4 &\\
\hline
\end{tabular}
\label{662table}
\caption{The gauge groups and matter content living on their world volume of each possible boundary state $\bf L_{I}$ is indicated. }
\end{table}

This spectrum is  obtained from Eqs. (\ref{cilindroshort}) and (\ref{moebius}).
For instance, we see that brane ${\bf L_{10}}= \,3\, 3 \,1 \,0^3$ is short, with $|\mathbf{S}|=6$. Thus, for the vector ($s_0=2$), a non vanishing contribution in (\ref{cilindroshort}) implies  $ \delta^{(2)}_{\sum_i \epsilon_i  \frac12(1-\psi \psi')} \delta_{\sum_i
\epsilon_i = 0}^{(2)} \ne 0$, namely $\psi =\psi'$. Moreover, for such choice of $\epsilon_i' s$ we see that  (\ref{moebius}) vanishes thus leading to the unitary group shown in the Table \ref{662table}. In a similar way, for the scalars ($s_0=0$) the states propagating between a boundary state and its orientifold image are selected, $\psi =-\psi'$. M\"obius amplitude (\ref{moebius}) is non vanishing in this case and produces a minus sign thus leading to antisymmetric representations.

The tadpole equations (\ref{tadod=6},\ref{taddd=6}) for this set of branes reads
\begin{eqnarray}
N_2+2N_3 + N_4 + N_5 +2N_6 +N_7 +2N_8 +N_9 +2N_{10}+N_{11}+3 N_{12}  =  16&&\\
N_1+2N_2 + 2N_3 + 2N_4 + N_5 +3N_6 +2N_8 +2N_9 +2N_{10}+2N_{11}+4 N_{12}  =  24&&
\label{662d6tadpoles}
\end{eqnarray}

States propagating between branes can be easily computed from (\ref{cilindro}) and (\ref{cilindroshort}). 
Two tensor multiplets are found in the internal sector (see for instance \cite{aaln}). It can be checked that all gauge and gravitational anomalies cancel.

At this point it may be instructive  and  useful for our
subsequent calculations to illustrate this in a detailed example containing only $({\bf L_1},{\bf L_6},{\bf L_{10}})$ states.

The  tadpole equations for the reduced set of D-branes lead to 
\beqa 
N_{10} &=&8-N_6=N1
\label{tadL1L6L10}
 \eeqa

The gauge group is $Sp( N_1)\times Sp( N_6)\times U( N_{10})$

with matter hypermultiplets in

\beqa && 3[(1,1,\Yasymm)+(1,1,{\bar \Yasymm})]+7(1,\Yasymm,1)+(1,\Ysymm,1)\\\nonumber
&+& ( N_1,
N_6,1)+(N_1,1,N_{10}) +(N_1,1,{\bar N_{10}})+
\\\nonumber && 3[(1, N_6,  N_{10})+(1, N_6, {\bar  N_{10}})]\eeqa

\begin{table}
\centering
\footnotesize
\[ {\renewcommand{\arraystretch}{1.3}
\begin{tabular}{|c|c|c|c|}
\hline
S-T & Internal & mult. & irrep.\\
\hline v & $(0,0)^6$                        & 1 & $Sp(N_1) \otimes
Sp(N_6) \otimes U( N_{10})$\\
s & $ \underline{(6,6)(2,2)}(0,0)(0,0)^3$ & 2 &
 $2(1,\Yasymm,1) + (1,1,\Yasymm)+(1,1,{\bar \Yasymm}) $\\
s & $ (4,4)^2(0,0)(0,0)^3 $             & 1 &
 $( N_1,N_6,1)+ 3(1,\Yasymm,1) +(1,\Ysymm,1)+ (1,1,\Yasymm)+ (1,1,{\bar\Yasymm})$\\
s & $(3,3)^2(1,1)(0,0)^3$ & 1 &
 $(1,N_6,N_{10})+(1,N_6,{\bar N}_{10}) + (N_1,1,N_{10}) + (N_1,1,{\bar N_{10}})$\\
s & $ {(5,5)(1,1)}(1,1)(0,0)^3 $& 1 &
 $2(1,N_6,N_{10})+2(1,N_6,{\bar N}_{10})$\\[.5ex]
\hline
\end{tabular}
}\]
\label{spectL1L6L10}
\caption{Massless spectrum of $6620^3$ example containing ${\bf L_1},{\bf L_6}$ and ${\bf L_{10}}$ boundary states (underlining indicates permutations}.
\end{table}

 It is easy to check that this spectrum (plus a closed sector containing two tensor multiplets and nineteen hypermultiplets) leads to vanishing of gauge and gravitational  anomalies if tadpole equations (\ref{tadL1L6L10}) are satisfied.





\section{Orbifolding ${( Gepner \,\, model)}^{c=6} \times \IT^{2}$}
Orbifolds of Gepner models are also easily implemented in the
language of boundary and crosscap states. The internal sector
described by ${( Gepner \,\, model)}^{c=6} \times \IT^{2}$ has
 a discrete symmetry acting on   fields in the following way

 \beq
g: Z \to e^{2\pi i v} \, Z ~~~~~  g: \Phi^{l_i}_{m_i s_i} \to
e^{2\pi i \frac{m_i \g_i}{k_i+2}}\Phi^{l_i}_{m_i s_i} \,\, i=1,\dots,r
\label{phasesymm}
 \eeq

where $Z=X^4 + i X^5$ denotes the complex coordinate on $\IT^{2}$ and
$(v; \g_i)$ are labels for  the generator $\hat g \in G$. For a torus with symmetry $\IZ_N$ we have $N v \in \IZ$. The labels $(v; \g_i)$ are conveniently encoded in terms of a simple current vector $j$

\beq j=(0, v; 2\g_1,\dots,2\g_r;0,\dots,0). \eeq

which satisfies $2\b_0 \bullet j \in \mathbf{Z}$ or in components

\beq
 -\frac{v}{2} + \sum \frac{\g_i}{k_i+2} = 0 \mod{\mathbf{Z}}.
\label{susycurrent}
\eeq

As it is well known,  twisted sectors must be included
in order to ensure the modular invariance of the torus partition function. As a consequence,  new tadpoles are expected to appear, in the transverse
channel, due to RR fields propagating in the twisted sectors.

The boundary states required  to cancel the tadpoles include the RR fields in the twisted sector of the theory.

When the  internal symmetry group is  $\IZ_N$, which is the case we are mainly interested in, we can  write an
expression for the boundary state in the simple case $v=0$ that  would correspond to    a four-dimensional compactification with $N=2$ supersymmetries.
The case  $v \ne 0$ will be considered later on in this section.

For a symmetry group $\IZ_N$ the twisted boundary states read
\beqa \nonumber\label{bound-orb} |\alpha\rangle_B&=&|S_0, S_{-1};(L_j,M_j,S_j)_{j=1}^r
\rangle_B=
         {\frac1{C_B}}
       \sum_{x=0}^{N}{\sum_{\lambda',\mu'}}^{\beta,b} (-1)^{\frac{s_0^2}2}\,
          e^{-i\pi\frac{s'_0\, S_0}{2}} e^{-i\pi\frac{s'_{-1}\, S_{-1}}{2}}\\ &&
          \prod_{j=1}^r \frac{S_{l'_j,L_j}}{\sqrt{S_{l'_j,0} }}\, e^{i\pi\frac{m'_j\, M_j}{k_j+2}}\,
      e^{-i\pi \frac{s'_j\,S_j}{ 2}} |\lambda',\mu'\rangle\rangle
\label{twbs}
      \eeqa

where now

\beq \label{carganula2} m_j = b + 2 \g_j x \eeq

with $x=0,1,\dots,N-1$ and $N$ is the order of the symmetry group
generated by the simple current $j$. The
branes are labelled as $\mathbf{L}, \mathbf{M}, S$ with
$M=(M_1,...,M_r)$ modulo group identifications and space-time labels $S_0, S_{-1}$ are defined in Appendix A . Short-orbit D-branes also include a $\psi$-label.

Replacing (\ref{carganula2}) into (\ref{bound-orb}) we see that the boundary state depends on $M_1,\dots,M_r$ only through
the phase

\beq
\label{fase}e^{\pi \frac{\sum M_i b}{k_i+2}} e^{2\pi i \frac{\sum \g_i M_i x}{k_i+2}}
\eeq

and therefore an independent set of labels for $|\a \rangle$ is given by
\beq M= H \sum_{i=1}^r \frac{M_i}{k_i+2} ~~~~ V = \sum_{i=1}^r
\frac{\g_i M_i}{k_i+2}. \eeq

In this way, $M$ 
represents nothing but   the action of the symmetry group
on the Chan-Paton factor.

In other words, if we begin with a configuration  of N coincident D-branes defined by the set
$\mathcal{M}=\{(M^{\a}_1,\dots,M^{\a}_r) ~~\a = 1,\dots,N \,|\, M = H \sum \frac{M^{\a}_i}{k_i+2}\} $, then the modding by $\G=(\g_1,...,\g_r)$ divides  $\mathcal{M}$ into classes
\beq
\mathcal{C}_I =\{ \sum{\frac{M_i^{\a}\g_i}{k_i+2}} := V_I  \},
\eeq
each with $N_I$ elements such that $\sum N_I = N$. From (\ref{fase}) the action on the Chan-Paton class $\mathcal{C}_I$
is given by the matrix $(\g)_{II} = e ^{2\pi i V_I}$ and the character of this representation

\beq
Tr \g^x = \sum N_I e^{2\pi i x V_I} .
\eeq


  Successive modding by  simple currents $j_2, j_3,...$
will introduce extra labels $W, X, \dots$, which at the end, if conveniently chosen, will be in one-to-one  correspondence with the labels $(M_1,\dots, M_r)$ 
of the A-type boundary states. This is expected because of the
Green/Plesser construction of the mirror theory relating  A-
and B-type models.

The spectrum of massless particles in the orbifolded theory is read  from the annulus amplitude. Given two boundary states
with   labels $\a =(\mathbf{L}, M, S)$ in class $I$ and $\tilde \a =( \mathbf{\tilde L}, \tilde M,\tilde  S) $ in class $J$,  the
amplitude between them in tree channel reads

\beq \label{cylin}
    Z_{\a I, \,\tilde \a J}^B(q)=\frac1C    \sum^{NS}_{\lambda'\mu'}
    \delta^{(H)}_{\frac{M-\tilde M}{ 2}+\sum{\frac{H}{ {2k_j+4}}m_j'}} e^{2 \pi x i{({V_I-\tilde V_J}}+\sum{\frac{\g_j m_j} {k_j+2}})}
    \prod_{j=1}^rN^{l_j'}_{L_j,\tilde L_j}
    \chi^{\lambda'}_{\mu'}(q) \eeq

 In this case, with the symmetry acting only on the Gepner sector, it is possible to  sum over  $x$  leading to the
 condition

 \beq
\label{proj}V_I-\tilde V_J+\sum{\frac{\g_j m_j} {k_j+2}} =0.
 \eeq

which implies that in general some states will be projected out of the original spectrum.


Under the orbifold projection the original full  cylinder amplitude changes as follows
\beq
N_{\a} N_{\tilde \a}Z_{\a \tilde \a} \to \sum_{x=0}^{N-1}\sum_{\a \tilde \a} Tr \g_{\a}^x \,Tr \tilde \g_{\tilde \a}^x  \,e^{2 \pi x i{\sum{\frac{\g_j m_j} {k_j+2}}} }Z_{\a \tilde \a}
\eeq

Interestingly enough, it is possible to rewrite the projection by
simple currents in such a way that its relation to the usual orbifolds
of toroidal manifolds is much more evident.  To see this we recall
that open string states formally read
\begin{equation}
|\Phi_k ; i,j\rangle \lambda^k_{ji}
\end{equation}
where $\lambda^k $ encodes the gauge group representation into which
the state $\Phi_k$ transforms. For instance, if the state $\Phi_0$
corresponds to gauge bosons, $ \lambda^0$ represents  gauge group
$G$ generators \footnote{ Which generically will be a product of
unitary, orthogonal and symplectic groups.}.

Let us assume that such Chan-Paton factors have been determined already and that we further act on string states with a
generator $\theta$ of a $Z_N$ symmetry group. Such action which
manifests as a phase $e^{2\pi i \frac{\g_i m_i}{k_i+2}}$ on world
sheet field $\Phi_k$ should, in principle, be accompanied by
corresponding representation of group action such that
\begin{eqnarray*}
\hat \theta|\Phi_k ; i,j\rangle \lambda_{ji}& = &\g_{ii'}|\hat \theta\Phi_k ; i',j'\rangle \g_{j'j}\lambda_{ji} \\
& = &e^{2\pi i \frac{\g_i
m_i}{k_i+2}}(\g^{-1}\lambda\g)_{j'i'}|\Phi_k ; i',j'\rangle
\end{eqnarray*}
Therefore, invariance under such action requires
\begin{equation}
e^{2\pi i \frac{\g_i m_i}{k_i+2}} \g^{-1}\lambda ^k \g = \lambda
^k \label{thetaprojection}
\end{equation}

By following the same steps as in Ref. \cite{afiv}, we can
represent $Z_N$ Chan-Paton twist in terms of Cartan generators as
$\g=e^{2\pi i VH}$ where $V$ is a ``shift"  eigenvalues vector of
the generic form
\begin{equation}
V=\frac{1}{N}(0^{N_0},1^{N_1},\dots,(N-1)^{N_{N-1}}) \label{twist}
\end{equation}
(ensuring $\g^N=1$) and Cartan generators are represented by $2
\times 2$ $\sigma_3$ submatrices.

On this basis, projection equation (\ref{thetaprojection}) reduces
to the simple condition
\begin{equation}
\rho _k V = \frac{\g_i m_i}{k_i+2} \label{proyeccion}
\end{equation}
where $\rho _k$ is the weight vector associated to the
corresponding $\lambda ^k$  representation.
This should be compared to (\ref{proj}).

In this latter framework the extension  to the case $v \ne 0$ is easily written down.
(\ref{proj}) is replaced with

 \beq
\label{proj2}V_I-\tilde V_J- \frac{v s_{-1}}{2} +\sum{\frac{\g_j m_j} {k_j+2}} =0.
 \eeq

where now $V_I $ represents the action on the Chan-Paton factor
due to the symmetry that acts simultaneously on the torus $\IT^2$ and the Gepner piece.

A last comment about the action on  Chan-Paton factors $\g_{I,J}$ is due. In the orientifold theory
we must introduce a boundary state and its image under $\Omega$. For long-orbit  D-branes $|B_{\mathbf{L},M}\rangle$
this yields an effective action

\beq
\g_{II} + \g_{II}^* = e^{2\pi i \frac{\sum \g_i M_i x}{k_i+2}}+e^{-2\pi i \frac{\sum \g_i M_i x}{k_i+2}}= 2 \cos{2\pi i \frac{\sum \g_i M_i x}{k_i+2}}
\eeq

which is real. This is simply the orientifold condition $\g_{\Omega} \g \g_{\Omega}^{-1} = \g^*$ which identifies
Chan-Paton factors.

For short-orbit D-Branes such that $\Omega \psi \to -\psi$, however, we have

\beq
\g_{II} |\alpha, + \rangle +\g_{II}^* |\alpha, - \rangle=
 e^{2\pi i \frac{\sum \g_i M_i x}{k_i+2}} |\alpha, + \rangle +e^{-2\pi i \frac{\sum \g_i M_i x}{k_i+2}} |\alpha, - \rangle
\eeq

Tadpole conditions can be generalized  for  orbifolded hybrid models $T^{2} \times Gepner$  in the following  way~\cite{aaj}
\begin{equation}
D_{\mu}^{\lambda}\left( {\Tr
\g_{N,2x}} +{\sqrt{\mathnormal{f}}} \Tr{\g_{D,2x,I}}\right) + O_{\mu}^{\lambda}    \cos \pi x v  =0 \label{tadhyb1}
\end{equation}

\begin{equation}
D_{\mu}^{\lambda}\left( {\Tr
\g_{N,2x+1}} +{\sqrt{\mathnormal{f}}} \Tr{\g_{D,2x+1,I}}\right)= 0
\label{tadhyb2}
\end{equation}

for all  states $(\lambda,\mu)$ such that $\chi^{\lambda}_{\mu+2 x \G}$ is  massless.




Here $O_{\mu}^{\lambda} $ is the orientifold charge we have in six dimensions for the state $(\lambda,\mu)$ while the factor ${f}=4 \sin^2{\pi x v}$ is a non-trivial contribution from the fixed points in the complex torus $\mathrm{T}^2$ 
in the NN sector. The labels N and D are used  to distinguish between D-branes with Neumann and Dirichlet boundary conditions
in the torus $T^2$.


\vspace{.5cm}
Before closing this section let us recap the general  steps to be followed in order to build up four dimensional models.
Our construction proceeds through  two consecutive stages. A first step is to build
a six dimensional model out of the possible Gepner models showed in Table \ref{tab:K3}. 
It is also necessary  to choose an orientifold projection as indicated in \ref{signos}. This gives rise to tadpoles
 which must be cancelled according to equations (\ref{tadorien}) and  (\ref{taddd=6}).
For each configuration of tadpole-canceling D-branes, the spectrum, the matter content and the gauge group, can be read
from (\ref{cilindro}), (\ref{cilindroshort}), ( \ref{moebius}) and (\ref{psi}). This completes the building
of  six-dimensional gauge theories.  
Further compactification
to four dimensions is achieved by  choosing an orbifold action on ${( Gepner \,\, model)}^{c=6} \times \IT^{2}$ as shown in (\ref{phasesymm}).
 Interestingly enough, spectra in the orbifold can be easily read using the simple expression (\ref{proyeccion}). Tadpole cancellation
conditions (\ref{tadhyb1}) and (\ref{tadhyb2}) will in general require the presence of additional D-branes with
Dirichlet boundary conditions on $T^2$ sitting at fixed points. The great advantage of this method is that 
six dimensional Gepner models are clearly easier to solve. If we are able to identify some of the distinctive features of the Standard Model in this first stage, say
the number of generations or the gauge group, then the steps down to four dimensions are quite direct and easy to implement.

\section{A MSSM example}

As an illustration of the general techniques discussed above we
concentrate here on a $\IZ_4$ modding of the   $[(6)^2(2)(0)]^{c=6}\times T^2$ model\footnote{We will write the internal sector in terms of four theories in what follows.}.
 Let us notice that inspection of allowed internal states
indicates that  only 3 massless chiral ($l_i=m_i, s_i=0$)  states, namely those such  $(m_1,m_2,m_3,m_4)=(3,3,1,0),\, (5,1,1,0),\, (1,5,1,0)$, do propagate
between brane ${\bf L_{10}}$ (with a $U(N_{10})$ gauge group living on its worldvolume) 
 and ${\bf L_6}$ (with an $Sp(N_6)$ gauge group).
Therefore, an internal modding of the form $\Gamma=(0,0,1,0)$
acting on the Gepner model will allow such states to remain in the
spectrum and,  by appropriately embedding it as a twist
$\gamma^{10}, \gamma^{6},\dots$ on the D-brane sector, the original
 $U(N_{10})\times Sp(N_6)$ gauge group could be broken
into a Standard-like model with 3 generations.
Moreover, in order to have ${\cal N}=1$ supersymmetry in four dimensions, we must accompany this modding with a $\IZ_2$ modding on $ T^2$, namely   $v_3=1/2$,  so as  to satisfy Eq.(\ref{susycurrent}).
Thus, our starting point is
\begin{equation}
\Gamma=(0,0,1,0)(\frac12)\otimes  \g^{a}
\label{gmodding}
\end{equation}

Note that the actual internal modding (see (\ref{phasesymm})) is  $\g_i/(k_i+2)$ so  it represents a $Z_4$ action.

As we stressed in the previous section, performing a $\IZ_2$ modding on the torus will require the introduction of a new set of branes having Dirichlet boundary conditions on the open string ends living on $T^2$. We quote them with an index $D$ while introducing an index $N$ to label the original branes with Neumann conditions on the third complex coordinate $Z$. We will refer to them as $D_Z$ and $N_Z$ branes respectively.

The generic tadpole equations (see Eq.(\ref{tadhyb1},\ref{tadhyb2}) for this model thus read

\beqa \label{t1}\sum_a D^a(\l,\mu) Tr \g_{0,N,a}+ O(\l,\mu)&=0&\\
\sum_a D^a(\l,\mu) Tr \g_{0,D,a}+ O(\l,\mu)&=0&\\
\sum_a D^a(\l,\mu) Tr \g_{2,N,a}&=0&\\
\sum_a D^a(\l,\mu) Tr \g_{2,D,a}&=0&\\
\sum_a D^a(\l,\mu) \left ( Tr \g_{1,N,a}+2Tr \g_{1,D,a} \right )&=0&
\label{gtadpoles}
 \eeqa

The indices indicate the order of the twist, the $D$ or $N$ sector on the torus, and the label $a$ for  a brane ${\bf L_a}$  (see Table \ref{662table}).
It is easy to see that any extra massless state is introduced in the closed sector by twisted internal states. 
Therefore, $D^a(\l,\mu)$ coefficients are just the coefficients appearing in the  untwisted tadpole equations (see (\ref{662d6tadpoles})) corresponding to the
vector $(2;0000;0)$ and the scalar state  $(0;2,2,2,0;0)$
\begin{eqnarray*}
N_2 + 2N_3 + 2N_4 + N_5 +2N_6 +N_7 +2N_8 +2N_9 +2N_{10}+N_{11}+3 N_{12} & = & 16\\
N_1+2N_2 + 2N_3 + 4N_4 + N_5 +3N_6 +2N_8 +4N_9 +2N_{10}+2N_{11}+4 N_{12} & = & 24
\end{eqnarray*}
and similarly for the $D_3$-branes sector.

As mentioned, ${\bf L_{10}}$ and ${\bf L_6}$ constitute the basic branes which, after splitting under modding action, will give rise to our model. 
It is interesting to remark that, both boundary states can  be placed on the same NN sector, or either ${\bf L_{10}}$ in NN sector and ${\bf L_6}$ in the DD sector (or viceversa)  or both in the DD sector. The basic features, discussed below,  will be independent of the sector choice. However phenomenological details will be different, mainly due  to the extra branes that must be  added to satisfy RR tadpole cancellation.

We choose two (minimal) $N_{10}=N_{6}=4$ stacks of  ${\bf L_6}$ and  ${\bf L_{10}}$ branes to start with a $U(4)\times Sp(4)$ gauge group. The modding  $\Gamma$ in ( \ref{gmodding}) is embedded as twists $\gamma_6$ and $\gamma_{10}$, on each respective stack, as

\begin{eqnarray}
\gamma_6& \rightarrow& V_6 =\uq(0,2) \\
\gamma_{10}&\rightarrow &V_{10}=\uq(1,1,1,3)
\end{eqnarray}

 For the vector $\Gamma.\mu =0$ and therefore from (\ref{proyeccion}) we find
\begin{eqnarray}
U(4) & \rightarrow& SU(3) \times U(1)^2  \\\nonumber
Sp(4)&\rightarrow &SU(2)\times SU(2)\nonumber
\label{groupbreaking}
\end{eqnarray}
where $Sp(2)\equiv SU(2)$. Thus, a LR symmetric-like  model group is obtained.

Moreover, the correct LR spectrum with 3 massless generations is found. Namely,   massless chiral states propagating in between
${\bf L_{10}}- {\bf L_{6}}$
\begin{eqnarray}
&&(0; 3  3  1  0;0 )\\\nonumber
&&(0; 1  5  1  0 0)\\\nonumber
&&( 0;5  1  1  0;0 )\nonumber
\label{3ciralstates}
\end{eqnarray}
satisfy
$\Gamma \mu=\frac14$ and therefore we find the  spectrum representations under
 $SU(3)\times SU(2)_L\times SU(2)_R \times Q_{B-L}$ to be
\begin{equation}
3[(3,2,1)_{\frac13}+(\bar3,1,2)_{-\frac13}+(1,1,2)_{1}+(1,2,1)_{-1}]
\end{equation}
where the subindex indicates the charge eigenvalue of
\begin{equation}
Q_{B-L}=\frac13 Q_3+Q
\end{equation}
 $Q_3$ being the generator of the  $U(1)$ in  $U(3)$ and $Q$ the
other  $U(1)$ generator in (\ref{groupbreaking}).

Actually, it is possible to establish a correspondence  with an intersecting brane model picture  in toroidal manifolds (see for instance  \cite{cim:2003} or \cite{cim:2002}).
Namely, under the action of $\gamma_{10}$ and $\gamma_{6}$,  boundary states ${\bf L_{10}}$ and ${\bf L_{6}}$ intersecting at a six dimensional manifold,  split into four stacks of boundary states  as
\begin{eqnarray}
{\bf L_{10}} [\,U(4)]&\rightarrow &{\bf L^a_{10}}[\,U(3)]+{\bf L^d_{10}}[\,U(1)]\\
{\bf L_{6}}[\,Sp(4)]& \rightarrow& {\bf L^b_{6}}[\,Sp(2)]+{\bf L^c_{6}}[\,Sp(2)]
\end{eqnarray}

where we have indicated in brackets the  gauge group living on the corresponding brane.
Thus, boundary states ${\bf L^a_{10}},{\bf L^d_{10}},{\bf  L^b_{6}},{\bf L^c_{6}}$ do match with the basic branes  $a,b,c,d$ arising in intersecting brane models on toroidal constructions (\cite{imr,cim:2002,Gmeiner:2005vz}).

\begin{figure}
\leavevmode \hskip 1.0cm \epsfxsize=15cm\epsfbox{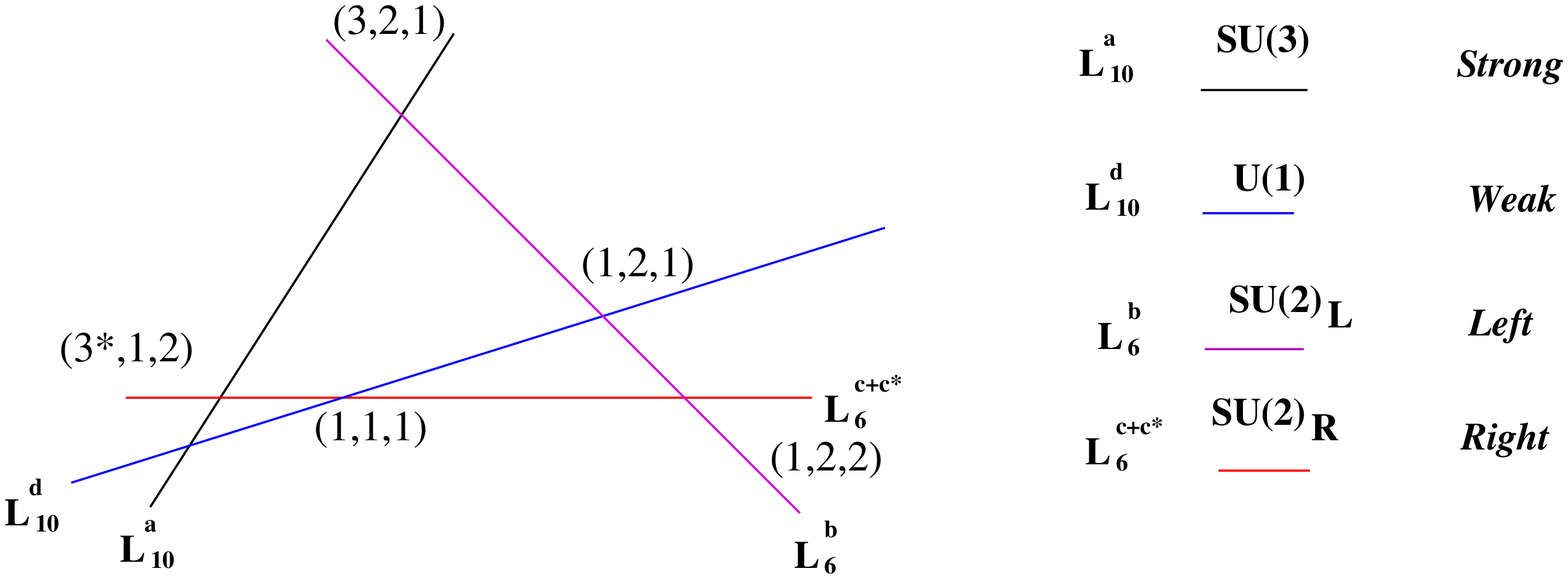}
\caption[]{ LR symmetric model obtained by orbifolding $6^2 2$ $D=6$ model. Original ${\bf L_{10}},{\bf L_{6}}$ boundary states do split, under the action of the modding, into ${\bf L^a_{10}},{\bf L^d_{10}}$ and ${\bf L^b_{6}},{\bf L^c_{6}}$ branes, giving rise to LR spectrum at the intersection.}
\label{LR}
\end{figure}

Thus, drawing boundary states as lines and interpreting multiplicities as intersection numbers we are lead to a graphic representation as the one given in  Figure \ref{LR}.

Besides states propagating between different branes we must consider states along the same type of branes.  They lead to vector like matter. 

Interestingly enough, massless states $( 4  0  2  0 ), ( 0  4  2  0 )$ and  $( 2  2  2  0 )$ do propagate in  ${\bf L_{6}}-{\bf L_{6}}$ sector. They satisfy $\Gamma.\mu=\frac12$  and thus, together with  $(1)( 0  0  0  0 )$, descending from  the six dimensional vector, lead to  
\begin{equation}
9(1,2,2)_0
\label{L6L6}
\end{equation}
candidates to LR  Higgses\footnote{They come from the seven antisymmetric and  one symmetric representations of $Sp(N_6)$ in Table(\ref{spectL1L6L10}) and the vector.}. There is also a pair of states $(1,2,1)_0+(1,1,2)_0$ descending from the symmetric representation. 

Notice that this sector is non chiral and that states fill up an ${\cal N}=2$ hypermultiplet

Branes $c$ and its image $c*$  are placed here on top of each other and on top of an orientifold point (leading to $Sp(2)$). Since such branes
 are parallel in the torus, following similar steps as discussed \cite{cim:2003}, we can think into separating them away from the orientifold point in the torus. Thus, $SU(2)_R \rightarrow U(1)_c$ where $U(1)_c$ charges are given by $T^3_R$ eigenvalues. Therfore, by introducing the weak hypercharge
\begin{equation}
Y=-T^3_R+\frac12 Q_{B-L}
\end{equation}
we find that the original LR symmetric model breaks down to $SU(3)\times SU(2)_L\times U(1)_Y$  MSSM with three chiral generations
\begin{equation}
3[(3,2,1)_{\frac16}+(\bar3,1)_{-\frac23}+(\bar3,1)_{\frac13}+
(1,1)_{\frac12}+(1,2)_{-\frac12}+(1,1)_{0}]
\end{equation}
including three right handed neutrini.
Moreover, LR chiral states  $(1,2,2)_0$ decompose into
$(1,2)_{-1/2}+(1,2)_{1/2}$ with the correct  MSSM Higgs charges.

\begin{figure}
\leavevmode \hskip 0.5cm \epsfxsize=15cm\epsfbox{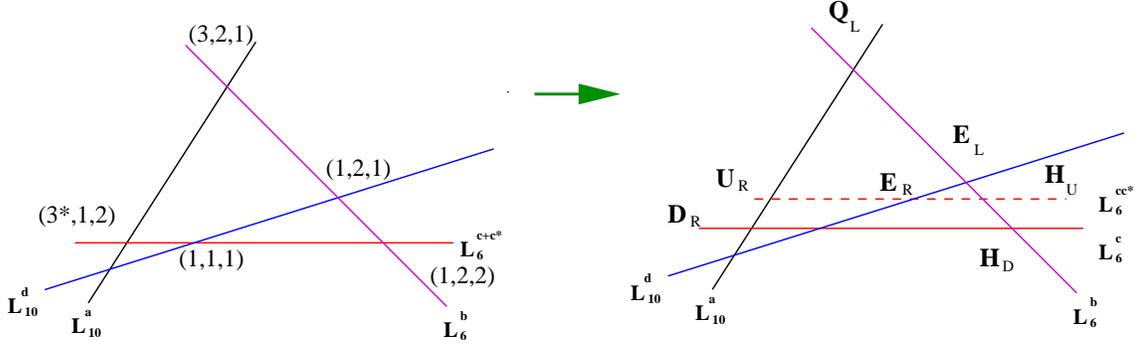}
\caption[]{By separating  ${\bf L^{c+c*}_{6}}\rightarrow {\bf L^c_{6}}+{\bf L^{c*}_{6}}$ away from the orientifold point breaking from LR to MSSM with 3 righthanded neutrini is achieved}
\label{LRSM}
\end{figure}

A pictorial representation is presented in Figure \ref{LRSM}.

Besides these basic boundary states leading to the  MSSM structure, additional stacks of branes must be added in order to satisfy tadpole cancellation equations. Different choices are possible  and each of them will give rise to particular phenomenological features.  Here we just want to show a simple choice that allows to complete the above construction to a fully consistent supersymmetric model.

With this aim we introduce a stack of $N_1$ ${\bf L_1}$  ``$N_Z$ branes"  and three stacks of $N'_5$ ${\bf L'_5}$,  $N'_1$ ${\bf L'_1}$, $N'_{10}$ ${\bf L'_{10}}$ ``$D_Z$-branes". Therefore the starting gauge group structure is
\beqa NN&:&SP(N_6)\times U(N_{10})\times SP(N_1)\\
DD&:&SP(N'_5)\times U(N'_{10})\times SP(N'_1)
 \eeqa

When performing the  modding given in Eq.(\ref{gmodding}) tadpole equations  (\ref{gtadpoles}) become

\begin{eqnarray*}
 2N_6 +2N_{10}& = & 16\\
N_1+ 3N_6  +2N_{10} & = & 24\\
2Tr\gamma^{2}_6 +2Tr\gamma^{2}_{10}& = & 0\\
Tr\gamma^{2}_1+ 3Tr\gamma^{2}_6  +2Tr\gamma^{2}_{10} & = & 0\\
 N'_5  +2N'_{10}& = & 16\\
N'_1+  N'_5  +2N'_{10} & = & 24\\
 Tr{\gamma'}^{2}_5  +2Tr{\gamma'}^{2}_{10}& = & 0\\
Tr{\gamma'^{2}}_1+  Tr{\gamma'}^{2}_5   +2Tr{\gamma'^{2}}_{10} & = & 0\\
2Tr{\gamma}_6+2 Tr{\gamma'}_5 +2 (Tr{\gamma}_{10}+2Tr{\gamma'}_{10})& = & 0\\
Tr{\gamma}_1+ 2 Tr{\gamma'}_1+  3 Tr{\gamma }_6  +2(Tr{\gamma}_{10}+2Tr{\gamma'}_{10}) & = & 0
\end{eqnarray*}

A solution is obtained by choosing $N_1=N_6=N_{10}=4$  in the DD sector and $N'_{10}=1$, $N'_{1}=8$, $N'_{5}=14$ with the corresponding  twists embedding (and induced gauge symmetry breaking) given in Table \ref{tabtwists}. Massless states propagating at intersection of different pairs of branes are shown in Table \ref{tabpnnspectrum} 

\bigskip
\bigskip
\begin{table}{}
\centering
\begin{minipage}{6.0in}
\footnotesize
\begin{tabular}{|l|l|l||l|}
\hline \hline Sector& Brane  &twist & Group\\
 \hline \hline NN&${\bf L_{10}}$ & $V_{10}=\frac14(1,1,1,3)$ &$U(4)\rightarrow U(3)\times U(1)$\\
{}&${\bf L_{6}}$ & $V_6=\frac14(0,2)$ &$Sp(4)\rightarrow  SU(2)_L\times SU(2)_R$\\
{}&$ {\bf L_{1}}$ &$V_1 = \frac14(1,1) $ &$Sp(4)\rightarrow  U(2)$\\
\hline \hline

 {DD}&${\bf L'_{10}}$ & $V'_{10}=\frac14(3)$&$U(1) $\\
 {}&$ {\bf L'_{1}}$&$V'_{1}=\frac14(0,2,1,1)$ &$SP(8)\rightarrow SP(2)\times \times SP(2)\times U(2)$\\
{}&$ {\bf L'_{5}}$&$V'_{5}=\frac14(0,0;2,2;1,1,1)$ &$SP(14)\rightarrow SP(4)\times SP(4)\times U(3)$\\
\hline\hline
\end{tabular}
\caption{Original ${\bf L_{I}}$ branes do split due to twist $V_{I}$. The original gauge group living on ${\bf L_{I}}$ world volume breaks accordingly.}
\label{tabtwists}
\end{minipage}
\end{table}

\vspace{1cm}
\begin{table}{}
\centering
\begin{minipage}{6.0 in}
\scriptsize
\begin{tabular}{|l|l|l|l|}
\hline \hline Sector& Branes  &States& IRREP\\
 \hline \hline NN&${\bf L_{10}}- {\bf L_{6}}$ &$(3310)$, $(5110)$,$(1510)$ &  SM\\
\hline N&$ {\bf L_{10}}- {\bf L_{1}}$ &$(3310)$ &none\\
\hline ND &${\bf L_{10}}- {\bf L'_{1}}$ &$(3310)$ &$(3,1,1;2,1,1)_{1/3}+(1,1,1;2,1,1)_{-1/3}+$
\\& & & $(3,1,1;1,2,1)_{-1/3}+
(1,1,1;1,2,1)_{1/3}$\\

\hline ND &${\bf L_{10}}- {\bf L'_{5}}$ &$(3310)$ &$(3,1,1;4,1,1)_{1/3}+(1,1,1;4,1,1)_{-1/3}+$
\\& & & $(3,1,1;1,4,1)_{-1/3}+
(1,1,1;1,4,1)_{1/3}$\\
\hline NN &${\bf L_{6}}- {\bf L_{1}}$ &$(4400)+(2200)$&none\\

\hline ND &${\bf L_{6}}- {\bf L'_{1}}$ &$(4400)+(2200)$
&$(1,{\underline {2,1}};{\underline {2,1}},1)_{0}$
\\ \hline
 ND &${\bf L_{6}}- {\bf L'_{10}}$ &$(3310), (5110),(1510)$
&none
\\ \hline
 \hline
 ND &${\bf L_{6}}- {\bf L'_{5}}$ &$(6200),(4400),(2220),(0420)$,
&$(1,{\underline {2,1}};{\underline {4,1}},1)_{0}$\\
\hline ND &${\bf L_{1}}- {\bf L'_{10}}$ &$(3310)$
&3(1,2,1;1)+3(1,1,2;1)\\
\hline ND &${\bf L_{1}}- {\bf L'_{5}}$ &$(2600),(4020)$
&$(2,1,3,1)+({\bar 2},1,{\bar 3},1)$\\
  & &
&$(2,1,{\bar 3},1)+({\bar 2},1,{ 3},1)$
\\

\hline DD &${\bf L'_{10}}- {\bf L'_{1}}$ &$(3310)$
&$(1;2,1,1)+(1;1,1,2)$\\

\hline DD &${\bf L'_{10}}- {\bf L'_{5}}$ &$(3310)$
&$(1,1,1;4,1,1)+(1,1,1;1,4,1)$

\\ \hline
\hline
\end{tabular}
\caption{Massless chiral primary states, denoted by $(m_1,\dots,m_6)$ propagating between a pair of branes ${\bf L_I}-{\bf L_J}$ are indicated. After performing the orbifold twist, the original gauge group in each brane breaks into several subgroups as in Table \ref{tabtwists} above. The last column shows the representations, with respect to such subgroups,  in which chiral superfields  accommodate.}
\label{tabpnnspectrum}
\end{minipage}
\end{table}

\bigskip
\bigskip
\newpage

A study of interactions among LR states and extra matter is beyond the scope of the present work. Nevertheless, some general remarks about Yukawa couplings can be advanced.

As a general observation notice that a Yukawa coupling will have the form
\begin{equation}
Y_{ijk}\Phi^i _{ba} \Phi^j_{ac} \Phi^k_{cb}
\end{equation}
where $\Phi _{ab}$ is the chiral superfield insertion connecting boundaries $a$ and $b$ and $i,j,k$ refers to internal CFT labels.  Such a term should be a singlet of the gauge group and invariant under $\Gamma$ modding. Moreover, it must be allowed by the fusion rules (\ref{su2fusion}) of the internal conformal field theory   \cite{dg,brunschom}.
Namely,

\begin{equation}
Y_{ijk}\propto \langle i j k\rangle\propto {\cal N}_{ij}^k
\end{equation}

For instance,  couplings like
 \begin{equation}
[( {\underline{ 5 1}} 1 0 )](3,2,1)_{1/3}^{ab} \times [( 3 3 1 0) )]({\bar 3},1,2)_{-1/3}^{ac}[( 2  2  2  0 )](1,2,2)_0^{bc}\
\end{equation}

(where we have indicated the internal charges in brackets) are   non vanishing and  lead to degenerate masses for two quark generations. Fusion rules forbid masses for the first quark generation (see (\ref{3ciralstates})).
A similar result is obtained for lepton masses since the same  internal states are involved for leptonic Yuakawa couplings.

The general pattern is very similar (the number of Higgses is different) to the one found in Ref.\cite{aiq} in the context of branes at singularities.

It is interesting to notice that  couplings of quarks or leptons to  states $[( {\underline{4  0}}  2  0 )](1,2,2)_0$ and 
$[(1)(0000)](1,2,2)_0$, discussed in (\ref{L6L6}), are not allowed by fusion rules. Thus, the model contains four effective LR Higgses.

In particular,  as addressed in in \cite{aiq}, the full picture of mass structures becomes more complicated   due, for instance, to the presence of  Yukawa couplings of quarks with colored triplets present at other intersections. For instance, D quarks  will couple to triplets in the ${\bf L_{10}}-{\bf L'_1}$ sector

\begin{equation}
[( 3 3 1 0) )]({\bar 3},1,2)_{-1/3}^{ac}\times
[( 3 3 1 0) )]({ 3},1,1;2,1,1)_{1/3}^{a1'}\times [( 2  2  0  0 )](1,1,2;2,1,1)_0^{1'c}\
\end{equation}
and therefore D quarks and triplets mix  once $SU(2)_R $ doublet acquires a vev. Through similar terms all the  three quarks would become massive.

Notice also that, from the 9  candidates to be interpreted as Higgs particles coming from ${\bf L_6}-{\bf L_6}$ sector, only those  with CFT quantum numbers $( 2  2  2  0 )$ are allowed in Yukawa couplings. For all of them, on the other hand, mass term like couplings are allowed. Thus, we can imagine a scenario where some of the  $(1,2,2)$ multiplets become very massive.

\subsubsection{An alternative with  the LR week sector on DD branes}

In the example discussed above the basic branes ${\bf L_{10}}$, containing strong group,  and ${\bf L_6}$, where  $SU(2)_L \times SU(2)_R$ lives, were placed in the same NN sector. However, it might be useful for future phenomenological applications, to place  the part of the spectrum
containing the $SU(2)_L \times SU(2)_R$ gauge theory  on the branes in the DD sector.

An interesting possibility of this kind is shown in Table \ref{tabL6DD}. In this case, even if ${\bf L_6}$ is placed in DD sector, tadpole cancellation requires to place a similar stack in NN sector thus  leading to two alternatives realizations of (3 generations) LR models.
Extra boundary states, required by consistency, are of the same kind we introduced in previous example, thus, states propagating between different pairs of branes can read directly from third column in Table [\ref{tabpnnspectrum}].
\begin{table}{}
\centering
\begin{minipage}{6.0in}
\footnotesize
\begin{tabular}{|l|l|l||l|}
\hline \hline Sector& Brane  &twist & Group\\
 \hline \hline NN&${\bf L_{10}}$ & $V_{10}=\frac14(1,1,1,3)$ &$U(4)\rightarrow U(3)\times U(1)$\\
{}&${\bf L_{6}}$ & $V_6=\frac14(0,2)$ &$Sp(4)\rightarrow  SU(2)_L\times SU(2)_R$\\
{}&$ {\bf L_{1}}$ &$V_1 = \frac14(1,1) $ &$Sp(4)\rightarrow  U(2)$\\
\hline \hline

 {DD}&${\bf L'_{10}}$ & $V'_{10}=\frac14(3)$&$U(1) $\\
{}&${\bf L'_{6}}$ & $V'_6=\frac14(0,2)$ &$Sp(4)\rightarrow  SU(2)_L\times SU(2)_R$\\
 {}&$ {\bf L'_{1}}$ &$V'_{1}=\frac14(1,1)$ & $Sp(4)\rightarrow  U(2)$\\
{}&$ {\bf L'_{5}}$&$V'_{5}=\frac14(1,1,1)$ &$SP(6)\rightarrow U(3)$\\
\hline\hline
\end{tabular}
\caption{An alternative construction leading to a  duplicated LR structure. The group 
$ SU(2)_L\times SU(2)_R$ can be chosen to be in the DD sector or in the NN sector.}
\label{tabL6DD}
\end{minipage}
\end{table}

It can be easily verified that this solution satisfies tadpole equations.

\subsubsection{Massless U(1) and K-theory constraints}

Anomalous $U(1)$ generators acquire mass through the Green-Schwarz mechanism. However, 
a non-anomalous U(1) may also become massive if there is an effective coupling
$ B\wedge F$. We must therefore ensure that  $Q_{B-L}$ is not one of them and remains massless.

For a  $U(1)_a$ gauge group on a brane  $a$, we will have the coupling 
\begin{equation}
\int _{[M_4]} (C_2^a- {C_2^a}')\wedge F^a 
\label{BF}
\end{equation}
where, in a geometrical setting,   $C_2^a$ (${C_2^a}'$ is its $\Om$-image)  come from the reduction of a $C_p$ form  on a supersymmetric cycle $a$ and $F_a$ is the $U(1)_a$ gauge field. 

Therefore, by expanding $ C_2^a$ forms, or analogously their corresponding cycles, into Ishibashi states, with expansion coefficients $D_a^i$ (and their $\Om$-images ${D'}_a^{i}$) (\cite{dhs0403}),  and requiring   $Q_{B-L}=\sum x_a Q^a$ coupling to vanish leads to

\begin{equation}
 N_a(D_a^i-{D_a^i}')x_a=0
\end{equation}

for each Ishibashi state $|i \rangle\rangle=|\lambda \mu ;x \rangle \rangle$ in the orbifold theory.

For the Ishibashi state $I=|(33)(33)(11)(00)^3; x=1 \rangle\rangle$  we obtain 
\begin{equation}
3 x_{a}i(D_{10,+}^I-D_{10,-}^I)- x_{d}i(D_{10,+}^I-D_{10,-}^I)=0
\end{equation}
where $D_{10,\pm}$ are the expansion coefficients for the parts $a$ or $d$ of the brane ${\bf L_{10}}$ (\ref{twbs}) in terms of Ishibashi states.

Then a solution with  $x_{a}=1/3$ and $x_{d}=1$
corresponds to having
\begin{equation}
Q_{B-L}=\frac13 Q_3+Q
\end{equation}
massless
. It can be shown that this is the only nontrivial condition.

Finally, there are additional constraints on the compactified theory coming from the fact that D-brane charges
are classified by K-theory \cite{Witten:1998cd}. One particular constraint is the vanishing of the Witten global anomaly which means that
 the number of massless fermions in the fundamental representation of a symplectic group is even. We have verified that
the Witten anomaly vanishes in the example we presented in the last section.

Generically, however, K-theory might impose  additional constraints.
  It would be interesting to further check consistency using maybe the method of probe-branes where additional constraints might appear ~\cite{U2000xp}. 

 \section{Summary and outlook}
In the first part of the present work we have addressed the construction of six dimensional Type IIB orientifold models based on a \textit{Gepner  models}  internal space  

Six dimensional models were constructed by considering stacks of B-type boundary states, required by a diagonal invariant partition function. Such boundary states  would correspond to D branes wrapping even cycles  of $K3$ \cite{eoty,ooy}. We have found the explicit  expressions for these boundary states and the rules to compute   their massless states spectra (associated  to open strings propagating among them). Tadpole  cancellation equations were also derived. 
Explicit computations  for the sixteen  diagonal Gepner models present in $D=6$ will be collected in \cite{aajd6gepners}. 

We have also shown how moddings by internal discrete symmetries and the so called parity and quantum dressings can be included in this context. In particular, A-type boundary models, corresponding to a charge conjugation invariant, should be obtainable \cite{greenplesser} by performing  possible moddings on B-type construction.

As shown in Ref.\cite{eoty,ooy},  more general boundary states, corresponding $SO(3)$ rotations, and including A and B-type cases, can be constructed in $K3$. It would be interesting to explore how such states could be obtained in the present context 

Following the ideas  presented in Ref.\cite{aaj},  four dimensional   chiral models were built by  further compactifying on a $T^2$ torus, sharing some of the symmetries of the $D=6$ models,  and by modding out by such symmetries.  The projection is realized as the combined action of a phase symmetry modding of the Gepner sector and a rotation of the torus lattice accompanied by a twist on Chan-Paton factors. The twist on Chan Paton factors can be viewed as a breaking of the original boundary states into component states with  specific  monodromy under the twist. Generically, when even order moddings are considered, new sets of branes are required for tadpole cancellation.

Interestingly enough, inspection of six dimensional spectra allows to identify phenomenologically  appealing models without the use of a computer scanning.

As an example of the construction, we described a $Z_4$ modding of the model $6620 \times \IT^2$. 
In such a model a basic structure of two stacks of four  boundary states , which we call ${\bf L}_{10}$ and ${\bf L}_{6}$, exist, with gauge groups $U(4)$ and $Sp(4)$  respectively, living on their  world volumes with three hypermultiplets propagating between them. These two stacks of boundary states  constitute the basic, six dimensional,  ``bulding blocks" of the MSSM. In fact, we showed that further compactification on a torus, accompanied by  $Z_4$ modding,  leads to the breaking 
\begin{eqnarray*}
{\bf L_{10}} [\,U(4)]&\rightarrow &{\bf L^a_{10}}[\,U(3)]+{\bf L^d_{10}}[\,U(1)]\\
{\bf L_{6}}[\,Sp(4)]& \rightarrow& {\bf L^b_{6}}[\,Sp(2)]+{\bf L^c_{6}}[\,Sp(2)]
\end{eqnarray*}
namely, into a four stacks of branes, giving rise to a Left-Right symmetric model with three massless generations living at the boundary states intersections.
Further  breaking to a MSSM (with the expected  three right handed neutrinos) can be achieved.

The four stacks $a,b,c,d$ of boundary states,  possess the ``basic building block"  properties used in intersecting brane realizations
\cite{imr,Marchesanothesis} of the Standard Model. They can be further embedded into a fully consistent supersymmetric orientifold model.

We have indicated  in the example some appealing features of the basic Yuakawa couplings structure. For instance,  the fact that vertex operators must connect different boundary states, the requirement of gauge and  discrete twist invariance and the CFT fusion rules allow to discard several terms. 
A detailed  investigation of  the structure of Yuakawa couplings  remains to be done. In particular, it would be interesting to see if a more systematic study, like in \cite{cim:2003} for intersecting branes, can be pursued in this context of RCFT.

An interesting feature of the hybrid construction is that lowering of the string scale \cite{Arkani-Hamed:1998rs,Antoniadis:1998ig, imr,CIMlower,Kokorelis:2002qi,Marchesanothesis} could be achieved by considering large extra  dimensions in the $T^2$ torus, transverse to the whole configuration  of intersecting boundary states.

 Indeed, in the present examples of the type  $Gepner \times T^2 /\IZ_N$,   boundary states would correspond to branes wrapping cycles on $K_3$ and stuck at a ${\bf C/\IZ_N}$ singularity.
Thus, if  we denote by  $V_4$ the volume of the Gepner piece, which should be of the order of the string scale  $V_4\propto 1/M_s^4$,  and by $V_2$ that of  two dimensional manifold. Then we expect the Planck scale, after dimensional reduction to four dimensions, to be
\begin{equation}
M_{Planck} \ =\  \frac2{\lambda }  M_s^4 \sqrt{V_4V_2} 
\end{equation}
where $\lambda$ is the string coupling. Therefore the string scale  $M_s$ can be lowered by choosing the volume $V_2$ ($V_2 \ =\  \frac{ M_{Planck}^2 \lambda ^2 }{4 M_s^4}$)  sufficiently large. Recall that the models constructed here are fully supersymmetric and though lowering the scale could be phenomenologically attractive in some cases it is not as compelling as in non supersymmetric models. 

Presumably, having these large extra dimension could allow for the introduction of dilute fluxes in a supergravity limit of some of these hybrid construction ~\cite{Grana:2005jc}.

\centerline{\bf Acknowledgments} We are grateful to A. Font, L.E. Iba\~nez and A. Uranga  
for stimulating  observations and
suggestions. 
G.A. work is partially supported by   ANPCyT grant PICT 11064 and CONICET PIP. J. J. is grateful 
to the Particle and Field Group of the Bariloche Atomic Center (CAB) for partial support and 
hospitality during the first part of this work.

\appendix
\section{SO(2d) Space- time partition functions}

We show the basic ingredients needed for  the computation of  modular transformation matrices of the space- time part of the partition functions of closed and open sectors. Even if we are mainly interested in $D=6$ dimensions we present here the general result in $D$ dimensions
Consider  $SO(2d)$ $d=\frac{D-2}2$ for dimensions $D=4,6,8$. There exist four representations $\Lambda>=|0>,|v>,|s>,|c>$ whose fundamental weights are encoded as
\begin{eqnarray}
|0>&=& (0,0,\dots,0)\\\nonumber
|v>&=& (1,0,\dots,0)\\\nonumber
|s> &=& (\frac12,\frac12,,\dots,\frac12)\\\nonumber
|c> &=& (\frac12,\frac12,,\dots,-\frac12)
\end{eqnarray}

Scalar product between two representations  $ \Lambda$ and $ \Lambda '$ is given by 
\begin{equation}
\Lambda.\Lambda'= \sum_{l=0}^{d-1} \Lambda_l\Lambda_l
\end{equation}

Recall that we need to redefine $\Lambda_{Gepner}=2\Lambda$ in order to have the normalization usually used in Gepner models.

The character that is associated to highest weight $\Lambda$,  at level one,  is given by $\chi_{\Lambda,1}(\tau)=\theta_{\Lambda,1}(\tau)$ \cite{goddardolive}. It leads to modular transformation matrices

\begin{eqnarray}
S_{\Lambda,\Lambda'}&=& e^{-2i\pi \Lambda.\Lambda' }\\
T_{\Lambda,\Lambda'}&=& e^{2i\pi (\Lambda^2-\frac{d}{24 })}\delta({\Lambda,\Lambda'})
\end{eqnarray}

Therefore, the space time matrix $P$ \cite{bs1} can be obtained as
\begin{equation}
\hat{P}_{\Lambda,\Lambda'}=T_{\Lambda}^{\frac12}S_{\Lambda,\Lambda''}T_{\Lambda'',\Lambda''}^2
S_{\Lambda'',\Lambda}T_{\Lambda''}^{\frac12}
\end{equation}

where $T_{\Lambda}^{\frac12}=e^{i\pi( h_{\Lambda}-\frac{d}{24 })}$ is the phase factor that is introduced in order to construct a real character from $\theta_{\Lambda,1}(\tau+\frac12)$ and  $h_{\Lambda}$ is the conformal weight. It coincides with $\Delta= \frac{\Lambda^2}{2}$ only in the case in which quantum numbers $\Lambda$ are given in the standard range above.
Thus, $P$ reads
\begin{equation}
{\hat P}_{\Lambda,\Lambda'}=e^{-i\pi\frac{d}{4 }}
e^{i\pi( h_{\Lambda}+h_{\Lambda'})}
\sum _{\Lambda''} e^{-2i\pi \Lambda.\Lambda'' }
e^{2i\pi (\Lambda'')^2 }e^{-2i\pi \Lambda''.\Lambda' }
\end{equation}

It is easy to see that, when all states are in the range above, the matrix $P$ is given by
\begin{equation}
{P}=\left(
\begin{array}{cccc}
c & s & 0 & 0 \\
s & -c & 0 & 0 \\
0 & 0 & \zeta c & i\zeta s \\
0 & 0 & i\zeta s & \zeta c
\end{array}
\right)
\end{equation}
with $c=cos\pi d/4$ and $s=sin\pi d/4$.

Since in the actual computation of M\"obius amplitude weights are shifted from the standard range by $\beta_i,\beta_0$ vectors, it appears useful to rewrite $P$ as (see for instance  (\cite{blumen0310} for $d=1$))

\begin{equation}
{\hat P}_{\Lambda,\Lambda'}=\sigma(\Lambda)\sigma(\Lambda')
e^{-i\pi\frac{d}{4 }}
e^{-i\pi {\Lambda  \Lambda'}
}
\sum _{\Lambda''}
e^{2i\pi (\Lambda''-\frac{\Lambda  +\Lambda'}2)^2 }
\label{pfinal}
\end{equation}
where

\begin{equation}
\sigma(\Lambda)=(-1)^{(\frac{\Lambda ^2}2 -h_{\Lambda})}
\end{equation}

and 
\begin{equation}
{\tilde P}_{\Lambda,\Lambda'}=
\sum _{\Lambda''}
e^{2i\pi (\Lambda''-\frac{\Lambda  +\Lambda'}2)^2 }
\label{ptilde}
\end{equation}
which, for NS weights (scalar or vector) reads

\begin{displaymath}
{\tilde P_{(\Lambda_0+\Lambda'_0 +\nu_i\beta_i )}^{NS}}=\left \{ \begin{array}{lll}
& 1 \quad  &d=1\\
&-e^{i\pi(\Lambda_0+\Lambda'_0 +\sum \nu_i)}\quad & d=3\\
& \frac12(1-e^{i\pi(\Lambda_0+\Lambda'_0 +\sum \nu_i )}&d=2
                         \end{array}\right.
\end{displaymath}

\section{The crosscap state in D=6 }

The $D=6$ spacetime bosons and fermions realize a $(2,2)$
superconformal algebra. The four world-sheet fermions
$\psi^{2,3,4,5}$ have an $SO(4)$ symmetry which requires them to be organized into
 unitary representations of the affine transverse Lorentz
algebra at level $k=1$. These are the scalar, vector, spinor and
conjugate spinor representations labelled respectively by $s_0=-1,
0, 1, 2$. It proves  convenient to split the representations of
$SO(4)$ at level $1$ into those of $SO(2)\times SO(2)$. The
latter are labelled by two numbers  $\Lambda_0, \Lambda_1 = -1, 0, 1, 2$
subject to $\Lambda_0+\Lambda_{1}=0 \mod{2\mathbf{Z}}$.

In order to implement a GSO projection we define the vectors
$\mu$'s and the inner product between them as
\begin{eqnarray*}
&& \lambda=(l_1, \dots, l_r)\\
 && \mu = (\Lambda_{0},\Lambda_1;s_1,\dots,s_r;m_1,\dots,m_r),\\
 && \mu\bullet\mu'\equiv-\frac{\Lambda_{0}\Lambda_{0}'}{4}-\frac{\Lambda_{1}\Lambda_{1}'}{4}
-\sum_{j}\frac{s_{j}s_{j}'}{4}+\sum_{j}\frac{m_jm_j'}{2(k_j+2)} .
\end{eqnarray*}
It is convenient to introduce  special vectors $\b_{0}$, $\b _{j}$
and $\tilde \b_{1}$
\begin{eqnarray*}
 && \b_{0}=(1,1;1,\dots,1;1,\dots,1),\\
 && \b_{j}=(0,2;0,\dots,0,\underset{j}{2}
,0,\dots,0;0,\dots,0),\ (j=1,\dots,r),\\
 && \tilde \b_{1}=(2,2;0,\dots,0;0,\dots,0).\\
\end{eqnarray*}
By using these vectors, we can construct the building blocks
 $\chi^{\lambda}_{\mu}(\tau)$ as
\begin{eqnarray*}
 &&\chi^{\lambda}_{\mu}(\tau) =
\chi_{\Lambda_{0}}(\tau)\chi_{\Lambda_{1}}(\tau)\chi_{m_1 s_1}^{l_1}(\tau)
\dots \chi_{m_r s_r}^{l_r}(\tau)
\end{eqnarray*}
where $\chi_{\Lambda_{0}}(\tau)$ and $\chi_{\Lambda_{1}}(\tau)$ are
$\widehat{SO(2)}_1$ characters. Then the GSO conditions and the
condition of fermionic sectors are
\begin{eqnarray}
 && 2\b_{0}\bullet \mu\in 2 \mathbf{Z}+1,\qquad
 \b_{j}\bullet \mu \in \mathbf{Z},\qquad
 \tilde \b_{1}\bullet \mu  \in \mathrm{Z}, \label{BetaCondition4}
\end{eqnarray}

The  type-B, GSO-projected partition function is then given by
  \beq
Z_D^{B}(\tau,\overline{\tau} )=\frac1{2^r} \frac{ ({\rm Im}\tau)^{-3}}{
|\eta(q)|^6 }
     \sum_{b_0=0}^{K-1} \sum_{\tilde b_{1}, b_1,\ldots,b_r=0}^1 {\sum_{\lambda,\mu}}^\beta
    (-1)^{s_0} \ \chi^\lambda_\mu (q)\,
    \chi^\lambda_{\mu+b_0\beta_0+\tilde b_{1} \tilde  \b_{1}
              +b_1 \beta_1 +\ldots b_r \, \beta_r} (\overline q) .\eeq
Here $K={\rm lcm}(4,2k_j+4)$ and ${\sum}^\beta$ means that the sum
is restricted to those $\lambda$ and $\mu$ satisfying
(\ref{BetaCondition4}).

The Klein bottle partition function  is obtained from
that of the torus by keeping states with equal left and right oscillators.  In
the direct channel it is given by

 \beqa \label{kb}Z_K^B &=
          &{\frac{4} {(4\pi^2 \alpha')^3}}\int_0^\infty  {\frac{dt} {4t^4}} {\rm Tr'}_{cl}\left(
             {\frac{\Omega} { 2}} e^{-4\pi t \left(L_0-{\frac{c} {24}}\right)} \right)
             \eeqa
where ${\rm Tr'}_{cl}$ denotes the trace over  the oscillator
modes in the closed string sector. The integration over the
bosonic zero modes yields the factor $(4\pi^2 \alpha')^3$.

The Klein bottle amplitude in the  transverse channel is obtained
by performing an $S$ modular transformation

\beqa
\widetilde{Z}_K^B &=&2^7  \prod_{j} \frac{\sqrt{k_j+2}}{2^{\frac{3r}2} K } \,
        \int_0^\infty  {dl}\,
      \frac1{\eta(2il)^3 }
     {\sum_{\lambda',\mu'}}^{ev} \sum_{\nu_0=0}^{K-1}
     \sum_{\nu_1,\ldots,\nu_r=0}^1
    \sum_{\epsilon_1,\ldots,\epsilon_r=0}^1\sum_{\eta_1,\ldots,\eta_r=0}^1
     \\\nonumber
&&
    (-1)^{\nu_0}\, \delta_{\Lambda_0',2+\nu_0
       +2\sum \nu_j}^{(4)}\delta_{\Lambda_1',\nu_0}^{(4)}
\\\nonumber
&&\prod_{j=1}^r \Biggl(  \frac{ P_{l'_j, \epsilon_j k_j} P_{l'_j,(\epsilon_j+\eta_j) k_j}}{
      S_{l'_j,0} }  \,
    \delta_{m_j',\nu_0+(1-\epsilon_j)(k_j+2)}^{(2k_j+4)}\,
      \delta_{s_j',\nu_0 +2 \nu_j+2(1-\epsilon_j)}^{(4)}
     \Biggr)
   \,   \chi^{\lambda'}_{\mu'} (2il)
\eeqa

where $l=1/t$.

From the Klein bottle amplitude in the transverse channel we can
read the expression for the crosscap state up to signs which are
contained in the M\"obius strip amplitude. The result is that the
crosscap state is given by

\begin{eqnarray}
 \label{crosscapbf2}
 | C\rangle^{NS}_B&=& 
\frac1{\kappa_c}
 {\sum_{\lambda',\mu'}}^{ev}  \sum_{\nu_0=0}^{\frac{K}2-1}
     \sum_{\nu_1,\ldots,\nu_r=0}^1  \sum_{\epsilon_1,\ldots,\epsilon_r=0}^1
   \,\,  \\ \nonumber
  & & \eta(\nu_0,\nu_i,\epsilon_j)
\delta_{\Lambda'_0,2+2\nu_0
     +2\sum \nu_j}^{(4)}\delta_{\Lambda'_{1},2\nu_0}^{(4)}
        \\\nonumber
 && \prod_{j=1}^r \Biggl( 
      \frac{P_{l'_j,\epsilon_j k_j}}{ \sqrt{S_{l'_j,0} }}  \,
   \delta_{m_j',2\nu_0+(1-\epsilon_j+\omega_j)(k_j+2)}^{(2k_j+4)}
 \delta_{s_j',2\nu_0 +2
\nu_j+2(1-\epsilon_j)}^{(4)} 
\Biggr) |\lambda',\mu'\rangle\rangle_c
\end{eqnarray}






We still have to fix the signs of the crosscap state.  As in
\cite{blumen0310}, the condition  that GSO orbits of
hatted  characters transforms,under the $P$-transformation, into themselves will be used as an ansatz to fix the signs in the crosscap state.

We want to compute the modular transformation of
\begin{equation}
{  M^\lambda_\mu=\sum_{\nu_0=0}^{\frac{K}2-1}
        \sum_{\nu_1,\ldots,\nu_r=0}^1 (-1)^{\left[ h^\lambda_{\mu}(\nu_0,\nu_j)-
         h^\lambda_{\mu}\right]}
    \,\, \widehat\chi^\lambda_{\mu+2\nu_0\beta_0+ \sum \nu_j \beta_j }
  (it+{\textstyle{\frac12}}) ,}
\label{msorbit}
\end{equation}

Thus, when we perform the $P$ transformation in (\ref{msorbit}) we
get

\begin{eqnarray}\nonumber
  M^\lambda_\mu &=&\sum _{\mu',\lambda'}^{\beta}
        \sum_{\nu_1,\ldots,\nu_r=0}^1
\prod_{k<l}(-1)^{\nu_l\nu_k}\sigma'_{(l',m',s')} e^{i\pi \sum
\nu_i(\Lambda_0+s_i-\Lambda'_0-s'_i+1)} e^{-i\pi{\mu.\mu'}}
\delta_{s_i+s_i',0}^{(1)} \\\nonumber
  & & \tilde P_{(\Lambda_0+\Lambda'_0 +\nu_i\beta_i )}
    \sum_{\epsilon=0}^1  P_{l,|\epsilon k-l'|}\, \delta_{m+m'+(1-\epsilon)(k+2)}^{(2)}\\&&
(-1)^{\epsilon_i (\frac{l_i'+m_i'}2+s_i')}(-1)^{\epsilon_i
(\frac{m_i}2+s_i+\nu_i)} \,\, \widehat\chi^\lambda_{\mu'}
(it+\frac12) \label{mstr3}
\end{eqnarray}

where the spacetime $\tilde P$-matrix is given in (\ref{ptilde}) for $d=2$.

By summing over $\nu_i$ we are lead to

\begin{eqnarray}\nonumber
  M^\lambda_\mu &=&\sum _{\mu',\lambda'}^{\beta}
        \sum_{\nu_1,\ldots,\nu_r=0}^1 \sigma'_{(l',m',s')}\\\nonumber
&& \prod_{k<l}(-1)^{\eta_l\eta_k} \delta^{(2)}_{\Lambda_0+\Lambda_1+\Lambda'_0+\Lambda'_1+\sum
\nu_i,0} 
e^{-i\pi{\mu.\mu'}} \delta_{s_i+s_i',0}^{(1)} \\\nonumber
  & &
    \sum_{\epsilon=0}^1  P_{l_i,|\epsilon k_i-l_i'|}\, \delta_{m_i+m_i'+(1-\epsilon_i)(k_i+2)}^{(2)}\\&&
(-1)^{\epsilon_i (\frac{l_i'+m_i'}2+s_i')}(-1)^{\epsilon_i (\frac{m_i}2+s_i)}
\,\, \widehat\chi^\lambda_{\mu'}
(it+\frac12)
\label{mstreta}
\end{eqnarray}

\begin{equation}
\eta_i=\Lambda_0 +s_i-\Lambda'_0-s'_i+1+\epsilon_i
\end{equation}

Using (\ref{mstreta}) we determine the unknown signs $\eta(\nu_0,\nu_i)$ in (\ref{crosscapbf2}) to obtain
the expression (\ref{crosscapbf}), for the crosscap state.


\section{ MS amplitude in the direct channel}

We present here the expression for the tree-channel M\"obius amplitude required to extract the gauge and matter field content. It can be computed from the amplitude of closed strings propagating between a boundary state (\ref{borde}) and
the crosscap state (\ref{crosscapbf})  and then performing a modular transformation to open string channel\footnote{ The corresponding
amplitude when the boundary state is short is essentially the same with a change 
in the normalization.}.

\begin{eqnarray}
\label{moebiloopdr}
   {{\cal Z}_M}^{B,NS}_{\alpha}&=&\langle C|q^{H}|{\alpha}\rangle _B =\\ \nonumber
&=& -\frac1{2^{r+1}}
         \int_0^\infty  \frac{dt}{t^5}\,
      \frac{1}{\eta(it+\frac12)^3 }
 {\sum_{\lambda,\mu}}^{ev}
     \sum_{\epsilon_1,\ldots,\epsilon_r=0}^1
     \left( \prod_{k<l} (-1)^{\rho_k\rho_l}\right)
   \\ \nonumber
     & &   e^{i(\pi/2) \sum \omega_j(m_j-2L_j +\epsilon_j(k_j+2))}
      (-1)^{\sum \omega_j \Lambda_0 /2}  \\ \nonumber
   & &   \delta_{\sum_j \rho_j, 1+ \frac{\Lambda_0+\Lambda_{1}}{2}+\sum \om_j}^{(2)}
    \, \delta_{\Lambda_0,0}^{(2)}\delta_{\Lambda_{1},0}^{(2)}
      \delta_{\sum \frac{K'}{ 2k_j+4}\left[ 2(M_j- \Delta_j)-m_j-
     \epsilon_j(k_j+2)\right],0}^{(K')}\\ \nonumber
   & &  \, \prod_{j=1}^r  \Biggl( \sigma_{(l_j,m_j,s_j)}\, \,
         Y_{L_j,\epsilon_j\, k_j}^{l_j}\, \,
       \delta_{m_j+\epsilon_j(k_j+2),0}^{(2)}\\\nonumber
     && \delta_{s_j,0}^{(2)}\,
            (-1)^{\frac{\epsilon_j}{2}\left[2S_j-s_j-2\epsilon_j\right]}\,
           (-1)^{\frac{(1-\epsilon_j)}{ 2}
              \left[2M_j-m_j-\epsilon_j(k_j+2)\right]} \Biggr)\\ \nonumber
     && \,   \widehat\chi^{\lambda}_{\mu} (it+{\textstyle\frac{1}{2}})
\end{eqnarray}
where $\rho_j= \frac{\L_0+s_j}{2} +\epsilon_j-1+\sum\om_j$.

Here
\begin{equation}
Y_{l_1,l_2}^{l_3}=\sum_{l=0}^k  \frac{S_{l_1,l}\, P_{l_2,l} \, P_{l_3,l}}{
                               S_{0,l} }=(-1)^{\frac{1-\epsilon_j}2(2L_j+lj)}N_{L_j,L_j}^{|\epsilon_jk_j-l_j|}
\end{equation} 
in terms   $SU(2)_k$ fusion coefficients \cite{difran}

\begin{displaymath}
N_{L_1,L_2}^{l}=\left \{ \begin{array}{lll}
&&1 \quad {\rm if} \quad |L_1-L_2|\le l\le {\rm min} \{{L_1+L_2,2k-(L_1+L_2)}\}\\
&& L_1+L_2+l= \,{\rm even }\\
&& N_{L_1,L_2}^{l}=0\, \quad  {\rm otherwise}
                         \end{array}\right.
\label{su2fusion}
\end{displaymath}

\end{document}